\newcommand\rh{\mbox{$r_{\mathrm{h}}$}}

\documentclass{aa}  
\usepackage[hidelinks]{hyperref}
\hypersetup{
colorlinks=true,
linkcolor=blue,
citecolor = blue,
urlcolor=cyan}
\usepackage{graphicx}
\usepackage{subcaption}
\usepackage{aadefs}
\usepackage{txfonts}
\usepackage{dblfloatfix}
\begin{document}

   \title{Ionic emissions and activity evolution in comet C/2020 F3 (NEOWISE): Insights from long-slit spectroscopy and photometry}

   \authorrunning{Aravind et al.}
    \titlerunning{Emissions in comet C/2020 F3}

    \author{K. Aravind\inst{1,2},
          E. Jehin\inst{1},
          S. Hmmidouch\inst{1,3},
          M. Vander Donckt\inst{1},
          S. Ganesh\inst{2},
          P. Rousselot\inst{4},
          P. Hardy\inst{4,5},
          D. Sahu\inst{6},
          J. Manfroid\inst{1} and 
          Z. Benkhaldoun\inst{3}
          }

   \institute{
   Space sciences, Technologies $\&$ Astrophysics Research (STAR) Institute, University of Liège, Liège, Belgium\\\email{aravind.krishnakumar@uliege.be}  \and Physical Research Laboratory, Ahmedabad-380009, India \and Cadi Ayyad University (UCA), Oukaimeden Observatory (OUCA), Faculté des Sciences Semlalia (FSSM), High Energy Physics, Astrophysics and Geoscience Laboratory (LPHEAG), Marrakech, Morocco \and Université Marie et Louis Pasteur, CNRS, Institut UTINAM (UMR 6213), OSU THETA, BP 1615, 25010 Besançon Cedex, France \and Université Bourgogne Europe, CNRS, Laboratoire Interdisciplinaire Carnot de Bourgogne ICB UMR 6303, 21000 Dijon, France \and
   Indian Institute of Astronomy, Bangalore-56034, India} 

   \date{Received ; }

  \abstract
   {} 
   {The long-period comet C/2020 F3 (NEOWISE) was the brightest comet in the northern hemisphere since C/1995 O1 (Hale-Bopp). Such comets offer a unique opportunity to study their composition and the spatial variation of the different emissions in detail. We conducted long-slit low-resolution spectroscopy and narrow-band photometry to track the evolution of its activity and composition during several weeks after perihelion. The images are used to compute the production rates of neutral molecular species and dust, while the spectrum is used to analyse the variation of emissions along the spatial axis in the sunward and anti-sunward directions to detect ionic emissions.}
   {Narrow-band (OH[3090 \AA], NH[3362 \AA], CN[3870 \AA], C$_2$[5140 \AA], C$_3$[4062 \AA], BC[4450 \AA], GC[5260 \AA], RC[7128 \AA]) and broad-band (Johnson-Cousins B, V, Rc, Ic) images of Comet C/2020 F3 were taken with TRAPPIST-North from 22 July to 10 September 2022 to track production rates, evolution of chemical mixing ratios with solar distance, and the proxy to the dust production (A(0)f$\rho$). A long-slit low-resolution spectrum was obtained on 24 July 2020 using HFOSC on the 2 m HCT at IAO, Hanle. Spectra extracted along the spatial axis in the sunward and antisunward directions enabled comparative analysis of the emissions in both directions.}
   {We report production rates and mixing ratios of OH, NH, CN, C$_2$, C$_3$, NH$_2$ and use forbidden oxygen line flux density to derive the water production rate. Ionic emissions from N$_2^+$, CO$^+$, CO$_2^+$ and H$_2$O$^+$, were detected at 4$\times$10$^4$ km to 1$\times$10$^5$ km from the photocentre in the tail direction. The average N$_2^+$/CO$^+$ ratio for the CO$^+$ (3-0) and (2-0) bands measured from the spectrum was $(3.0\pm1.0)\times10^{-2}$, further refined to $(4.8\pm2.4)\times10^{-2}$ using fluorescence modelling techniques. We measure CO$_2^+$/CO$^+$ ratio to be 1.34$\pm$0.21. Combining both the N$_2^+$/CO$^+$ and CO$_2^+$/CO$^+$ ratios, we suggest the comet to have formed in the cold mid-outer nebula ($\sim$50~K-70~K). Furthermore, the average rotation period of the comet was calculated to be $7.28\pm0.79$ hours with a CN gas outflow velocity of $2.40\pm0.25$ km/s.}
   {}


   \maketitle
%

\section{Introduction}\label{introduction}
Comets, celestial bodies composed of ice, dust, and organic compounds, offer invaluable insights into the early stages of our Solar System's formation and evolution. They serve as time capsules, preserving clues about the conditions and materials present during the Solar System's infancy. Consequently, the examination of comets presents a distinctive opportunity to explore the physical and chemical processes that took place during the nascent stages of the formation and evolution of our Solar System.

One of the most intriguing aspects of comets is their dynamic nature, especially during close approaches to the Sun. Comet C/2020 F3 (NEOWISE), hereafter 20F3, is a dynamically old long-period comet\footnote{\url{https://www.oaa.gr.jp/~oaacs/nk/nk4202.htm}} with a near-parabolic orbit discovered on 27 March 2020 by the NEOWISE space telescope \citep{NEOWISE} at magnitude 18.0 when it was at 2.00 au from the Sun. 20F3 was the brightest comet in the northern hemisphere since the comet Hale-Bopp in 1997. The comet reached maximum brightness in early July 2020 with a magnitude of 1, making it bright enough to be visible to the naked eye. The comet with an orbital inclination of 128.92$^\circ$ reached its perihelion on 3 July 2020, at 0.29 au, and its closest approach to Earth occurred on 23 July at a distance of 0.69 au. Strong sodium doublet emission at 5890 {\AA} was detected at different epochs in July \citep{Lin2020ATel,Cochran2020DPS}. H$_2$O$^+$ lines were detected at the visible red wavelength (5800 to 7400 {\AA}), although they are much weaker than the sodium line, suggesting that the straight red tail reported by numerous observers in mid-July was probably dominated by sodium atoms rather than H$_2$O$^+$ ions \citep{Ye2020DPS}. In support of this, the image\footnote{\url{https://astron-soc.in/outreach/wp-content/uploads/2020/Neowise2020/ContestWinners/HM1_33.jpg}} mentioned in \cite{atel_aravind}, taken on 24 July 2020 at the Indian Astronomical Observatory (IAO) site by Dorje Angchuk, clearly illustrates a distinctive blue ion tail. Taking into account that the image was taken while the comet was setting, the strong ion tail is seen to be aligned in the eastward direction. 

Comets exhibiting extensive plasma tails as they approach the Sun closely offer a unique opportunity to study their composition and behaviour. Investigating the relative abundances of key ionic species as a function of the radial distance from the nucleus is one of the best ways to understand the formation and interaction of cometary material with the solar wind \citep{N2plus_H2Op_g_lutz}. Such studies can provide essential insights into the physical and chemical processes occurring within the comet's coma and tail.

Although detection of major neutral gas species such as CN, C$_2$, C$_3$, and NH$_2$ in the visible range is common, fewer observations have reported the detection of emissions from molecular ions such as N$_2^+$, CO$^+$, and H$_2$O$^+$ \citep[eg., ][ and references therein.]{N2plus_H2Op_g_lutz, wyckoff_Halley_ions, wyckoff_ionsincometails, cochran_N2+_CO+_many, jockers_ionsHalley_CO2+, Kawakita_ions_unidentified, Cochran_ions_N2+_CO2+, Biver_C2016R2, Cochran_C2016R2, Korsun_C2002VQ94, Umbach_ions, Opitom_R2_highres, Kumar_R2}. Furthermore, the detection of emissions from CO$_2^+$ has been reported for a few comets like Bradfield (1979 X) and Seargent (1978 XV) \citep{bradfield_seargent, weaver}, 1P/Halley \citep{Feldman_CO2p,wyckoff_Halley_ions, jockers_ionsHalley_CO2+, Umbach_ions}, C/2002 C1 (IKEYA-ZHANG) \citep{Cochran_ions_N2+_CO2+} and C/2016 R2 (PanSTARRS) \citep{Opitom_R2_highres}. Most of these detections mentioned above were made in the near-ultraviolet (NUV) regime, and only a handful of reports are for CO$_2^+$ detections in the optical wavelength range.

In this paper, we use long-slit, low-resolution optical spectroscopy to calculate the production rates of neutral molecular emissions and report the detection of ionic species—N$_2^+$, CO$^+$, CO$_2^+$, and H$_2$O$^+$—in comet 20F3, providing new insights into its compositional characteristics. We compare these results with those of radicals observed in the optical spectral range during the same period using the TRAPPIST-North Telescope. Following the introduction in Section \ref{introduction}, Section \ref{Observations} details our observations and the associated data reduction techniques. The evolution of activity and composition, including production rates and dust proxies, along with the detection of ionic emissions, is elaborated in Section \ref{discussion}.

\section{Observations and data reduction}\label{Observations}
\subsection{Photometry (TRAPPIST)}

We used the TRAPPIST-North telescope to observe the activity of the bright comet 20F3 post-perihelion for 18 nights, from 22 July (r$_h$ = 0.63 au, outbound) to 10 September (r$_h$ = 1.61 au, outbound) 2020 obtaining a total of 399 images in various filters. The comet was very bright when we started the observations, with a visual magnitude of about 4.0, and also pretty low on the horizon. We had many consecutive good nights, and we could follow the drop in activity of the comet day after day before it was no longer visible.

TRAPPIST, short for TRAnsiting Planets and PlanetesImals Small Telescope, consists of two 60-cm robotic telescopes dedicated to detecting and characterising exoplanets via the transit method, as well as studying small Solar System bodies such as asteroids and comets \citep{Jehin2011}. TRAPPIST-North (TN, Z53), was installed in 2016 at the Oukaimeden Observatory in Morocco's Atlas Mountains in collaboration with the Cadi Ayyad University. It is a Ritchey-Chretien telescope with a focal ratio of F/8, mounted on a direct drive German equatorial mount. TN is equipped with a thermo-electrically cooled deep depletion 2K$\times$2K Andor IKONL CCD camera, offering a 20$^\prime \times$20$^\prime$ field of view with a high sensitivity. The pixel binning (2x2) was applied, resulting in a plate scale of 1.2$^{\prime\prime}$/pixel. We tracked the evolution of gas production rates of the OH, NH, CN, C$_3$, and C$_2$ species using HB narrowband filters \citep{Ahearn_85, Farnham2000, Moulane2018}. Furthermore, we measured both the magnitude and the A(0)f$\rho$ parameter, a proxy to the dust production \citep{A'Hearn1984} with broad-band filters (B, V, Rc and Ic) and the blue, green and red narrow-band dust continuum filters (BC, GC, RC) \citep{Farnham2000}.

Data calibration was carried out using standard procedures, with regularly updated master bias, flat, and dark frames. Sky contamination was removed, and flux calibration was achieved using frequently updated zero points. To determine the gas production rates from the HB narrow-band images, we computed the median and radial brightness profiles for both the gas filter images and the BC images within an annulus aperture between nucleocentric distances of $10^{3.6}$ and $10^{4.1}$ km. The latter is less affected by cometary gas emissions than other dust filters \citep{Farnham2000}. This allows us to effectively subtract dust contamination from the gas filters using the adjusted flux in the uncontaminated BC filter, using the dust colour in narrow band filters (BC-GC), following the procedure described in \cite{Farnham2000}. The surface brightness was then converted to column density, and the resulting profiles were fitted using a Haser model \citep{haser, bodewits_haser}, assuming an outflow velocity of 1 km/s and effective scale lengths from \citep{Ahearn_85}. Further details on this data reduction process can be found in \citep{Moulane2018}, where the same methodology was applied. The Haser model fit was performed within the annulus mentioned above to decrease the aperture effect and lessen dust contamination, and to minimise the effects of PSF and atmosphere near the optocenter of the comet. The dust contribution is less extended than the gas and peaks at the nucleus. Beyond this distance, signals, particularly in the OH filter, tend to diminish as a result of the lower signal-to-noise ratio. Fluorescence efficiencies, or g-factors as defined in \cite{Ahearn_85}, were taken from Schleicher's website\footnote{\url{https://asteroid.lowell.edu/comet/gfactor.html}} for the corresponding heliocentric distances, to convert the photon density into column densities. Furthermore, we calculated the Af$\rho$ parameter within 10,000 km of the nucleus and applied a phase angle correction using the phase function normalised at $\theta$ = 0$^\circ$, based on the composite phase function described by David Schleicher\footnote{\url{https://asteroid.lowell.edu/comet/dustphase.html}}, which incorporates elements from \citealp{Schleicher1998} and \citealp{Marcus2007}.
  \begin{figure}[h!]
  \centering
   \includegraphics[width=0.6\linewidth]{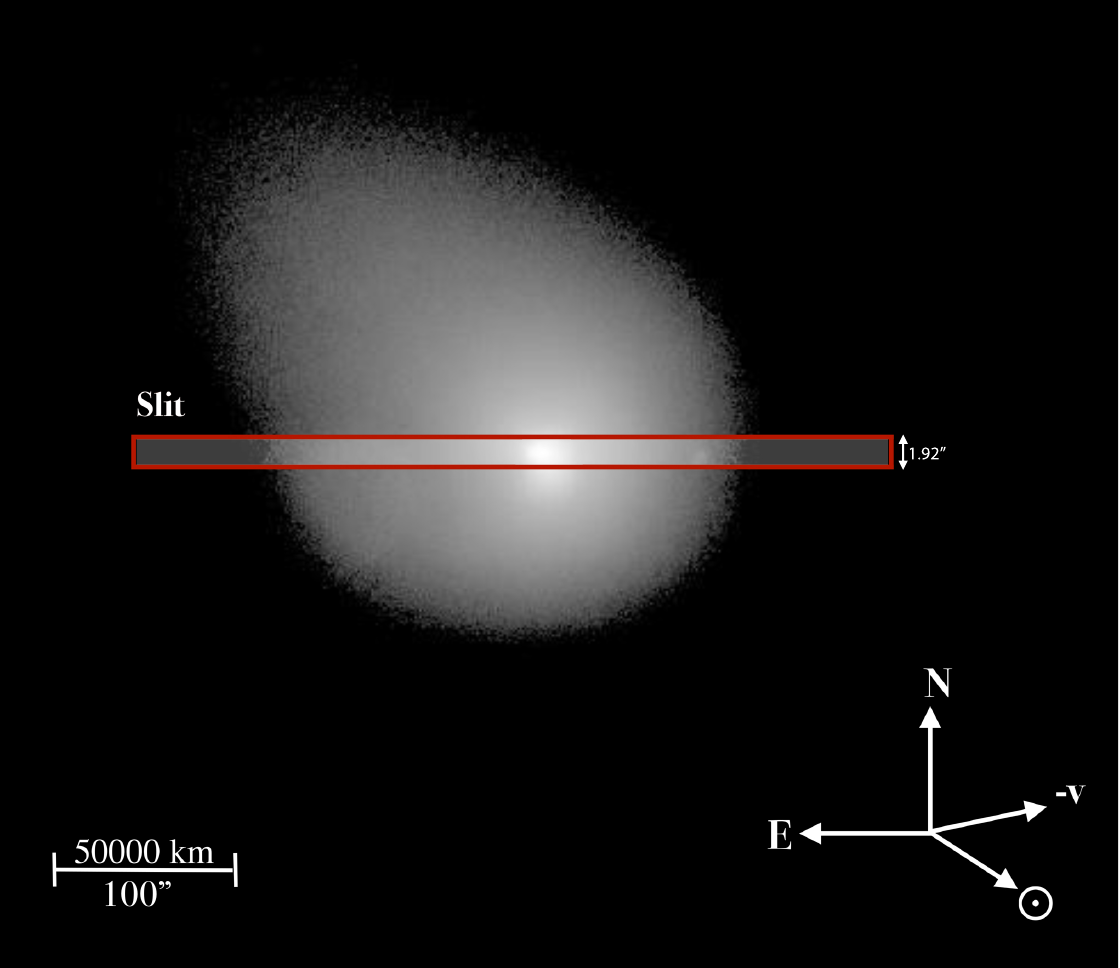}
      \caption{The orientations of the slit overlaid on an HFOSC image at the time of acquisition on 24 July 2020.
              }
         \label{F3_slit}
   \end{figure}

\subsection{Long slit spectroscopy (HCT)}\label{longslit}
The spectroscopic observations were carried out with the Hanle Faint Object Spectrograph and Camera (HFOSC) mounted on the 2-m Himalayan Chandra Telescope (HCT) at the Indian Astronomical Observatory \citep[see][for further details about the use of the HCT and HFOSC for comet observations]{156P_aravind}. A 11\arcmin~long and 1.92\arcsec~wide slit, corresponding to a resolving power of $\sim$1300 (providing a spectral resolution of $\sim$1.45\AA/pixel), was used for the comet observations. The observations were carried out on the night of 24 July 2020 when the comet was at a heliocentric distance of 0.66 au and a geocentric distance of 0.69 au.
   
Two spectra of 300 seconds each were obtained with the comet at the centre of the slit oriented along the east-west direction (see Figure \ref{F3_slit}). Spectroscopic standard star Feige 110 was also observed, with a wider slit (15.4\arcsec) to avoid light loss, to construct the sensitivity function of the instrument required for the flux calibration. The halogen lamp spectra, zero exposure frames, and FeAr spectra were also obtained for flat fielding, bias subtraction, and wavelength calibration, respectively.

The comet spectroscopic data were reduced and calibrated using self-scripted Python codes and IRAF, following normal long-slit procedures as detailed in \cite{156P_aravind}. Due to the time constraint in obtaining a separate sky frame, the sky spectra were extracted for similar apertures from a standard star frame observed for similar exposure time and similar airmass on the same night. The corresponding sky spectra were used for effective sky subtraction from the comet spectra.

\begin{figure}[h!]
  \centering
   \includegraphics[width=0.82\linewidth]{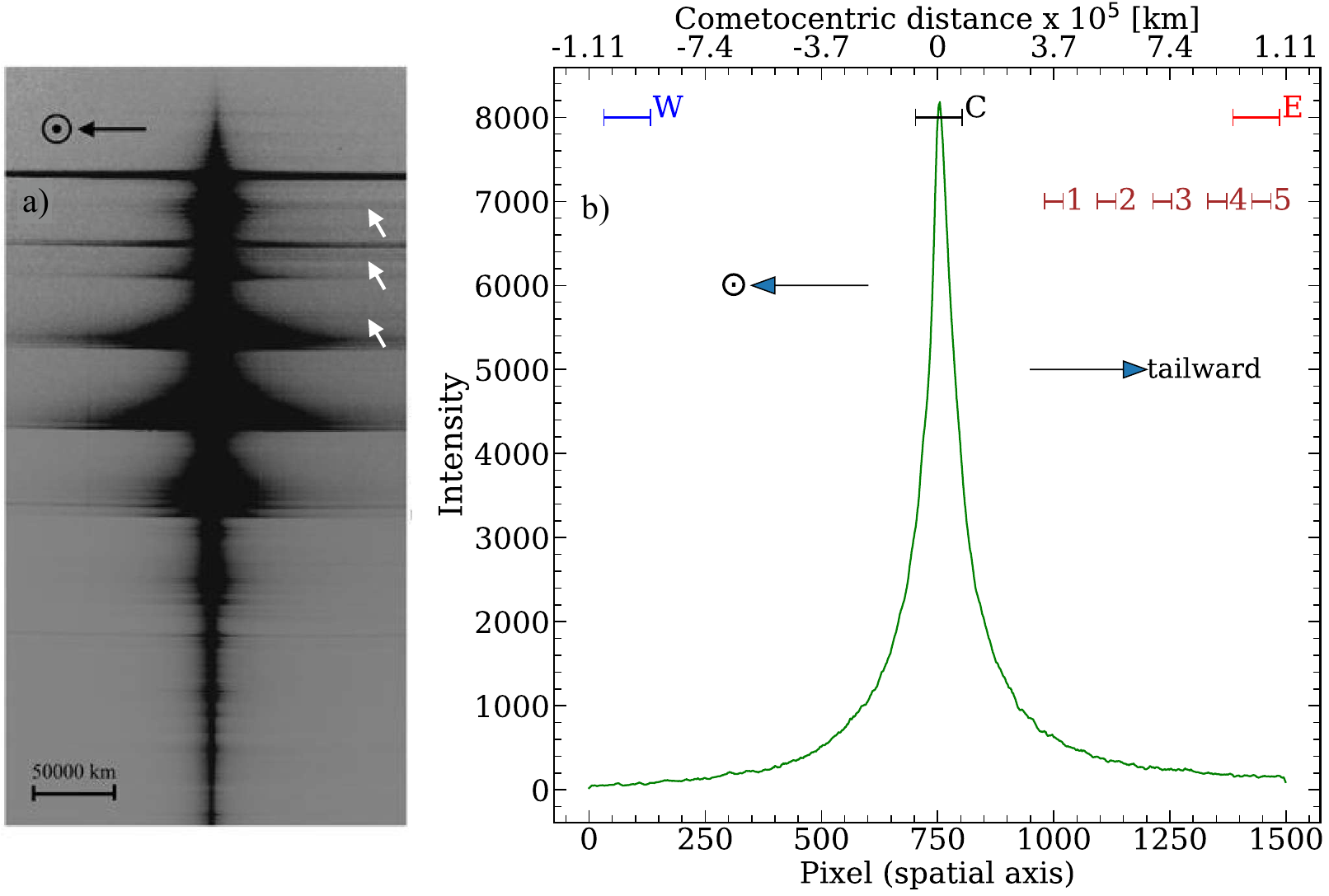}
      \caption{Illustration of the (a)  2D spectrum of 20F3 with the blue wavelength starting from the top side. White arrows denote the visually noted doublets in the anti-sunward direction; (b) the various apertures (horizontal bars) chosen for the spectral extraction compared to the comet photon density profile around the nucleus. E and W represent the apertures used in the East and West directions, respectively.
              }
         \label{F3_aperture}
   \end{figure} 
   
      \begin{figure}[h!]  
   \includegraphics[width=0.95\linewidth]{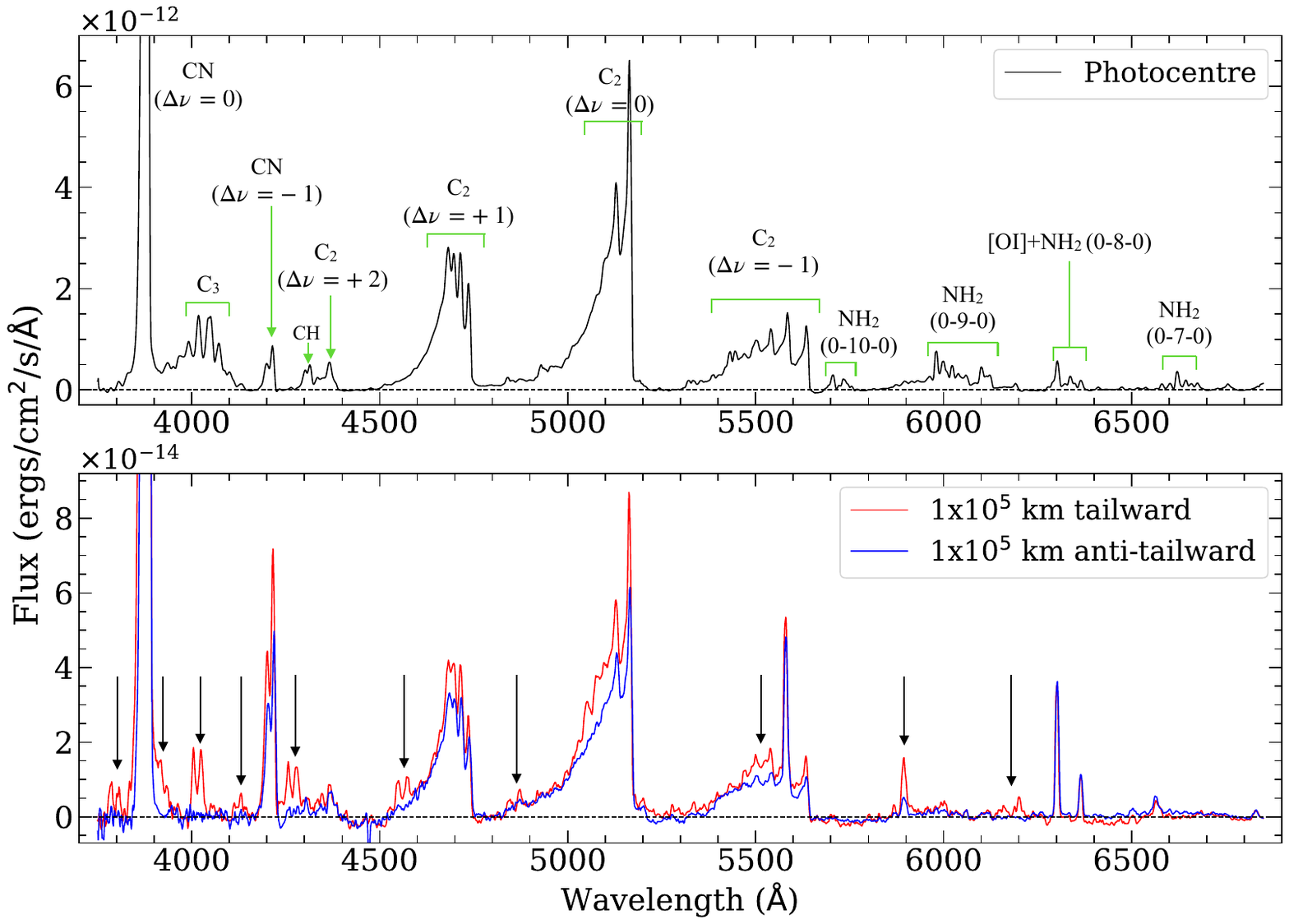}
      \caption{Long slit spectra of 20F3 observed on 2020 June 24, extracted at photocentre (top panel) and at the extreme regions of the slits in the tailward and anti-tailward directions (bottom panel). The arrows depict the regions of dissimilarity for spectra extracted from extreme regions of the slit.  }
         \label{F3_centre_ions}
   \end{figure}

   \begin{figure}[h!]
	\centering	\includegraphics[width=0.7\linewidth]{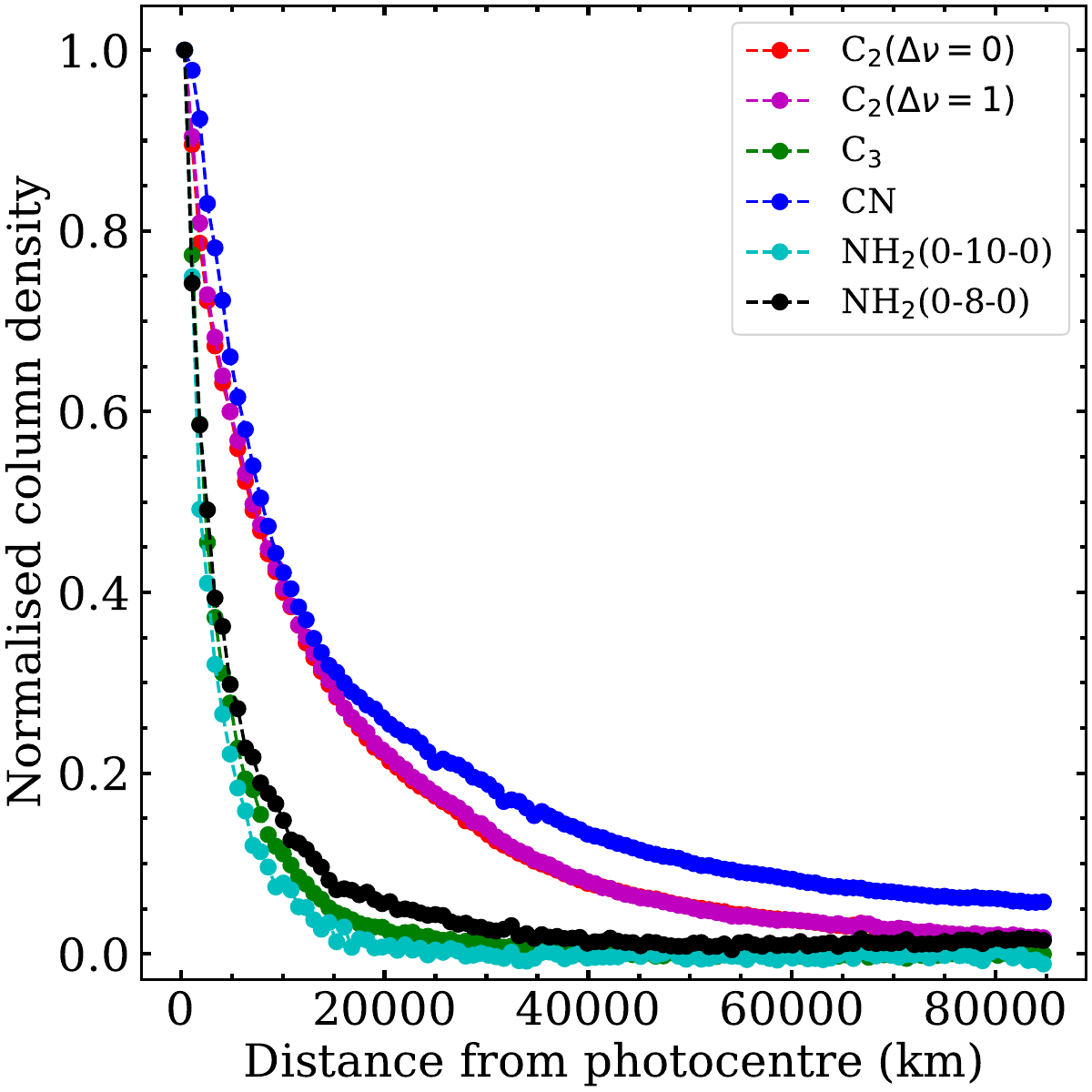}
	\caption{Median column density profiles of the major molecules detected in the spectrum.} 
	\label{F3_CD}
\end{figure}

\begin{figure*}[h!]
	\centering	\includegraphics[width=0.7\linewidth]{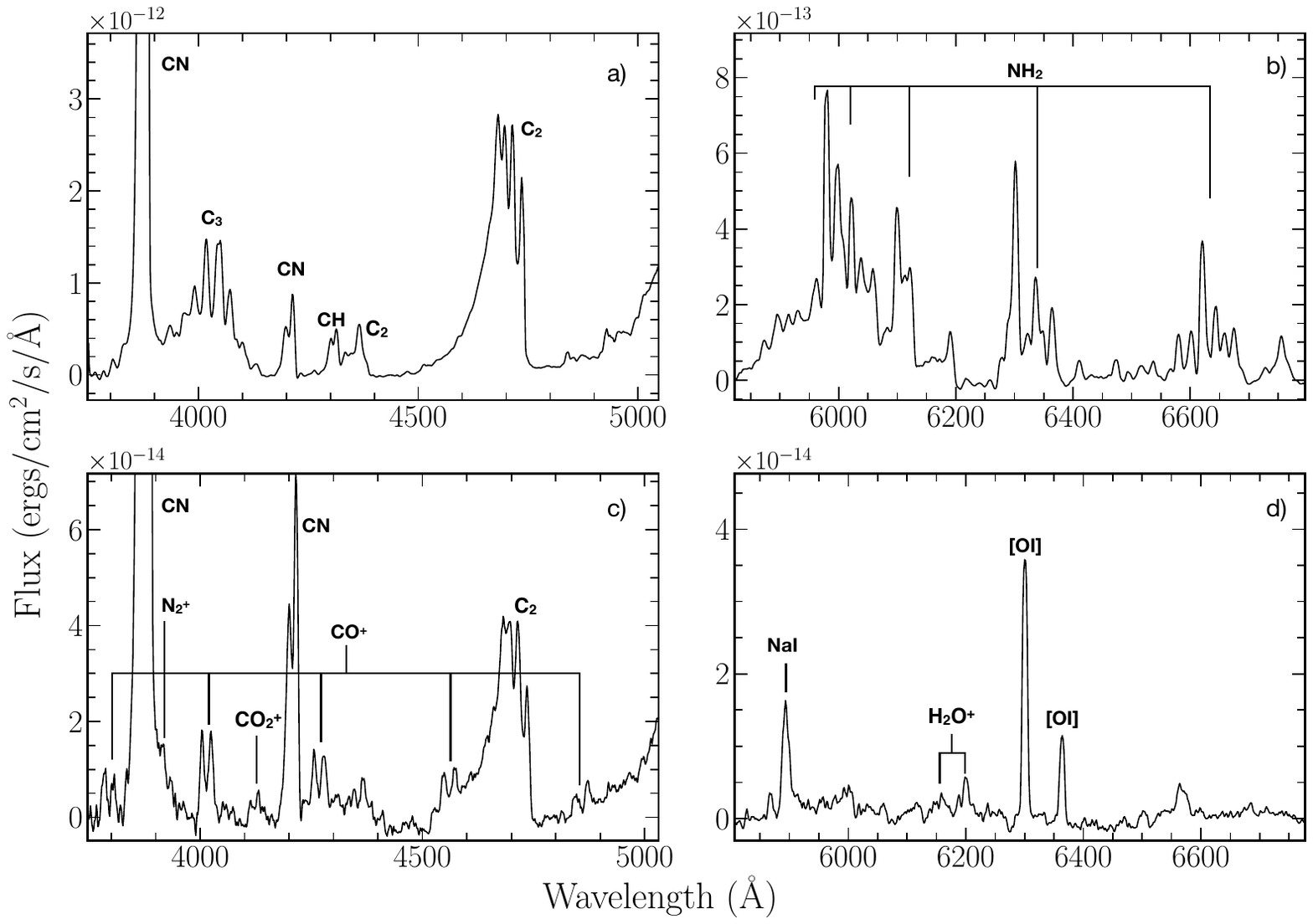}
	\caption{Depiction of detected emissions for spectra extracted (a) and (b) at photocentre; (c) and (d) 10$^5$ km in the tailward direction.} 
	\label{F3_tail}
\end{figure*}
From the preliminary analysis of the 2D spectra of 20F3, as shown in panel (a) of Figure \ref{F3_aperture}, the appearance of doublet emissions at certain positions in the anti-sunward direction was noted. At the time of observation, the slit and the anti-Sunward direction only had an angular difference of 22$^\circ$ in the clockwise direction. Hence, considering the orientation of the ion tail and the possibility of coincidental projection orientation with the slit direction (eastward, see Figure \ref{F3_slit}), the presence of ionic emissions was explored by analysing the emissions in spatial directions, both westward (anti-tailward / sunward) and eastward (tailward / anti-sunward). For this reason, three cometary spectra, at the photocentre (C) and at the two ends of the slit (202\arcsec away from the photocentre), eastwards (E) and westwards (W), each for an aperture of 30\arcsec~were extracted from the observed long-slit spectra corresponding to the locations C, E and W, respectively, as shown in panel (b) of Figure \ref{F3_aperture}.

The comet spectrum is comprised of both the coma gas emission spectrum and the spectrum of the sunlight scattered by the dust particles. Hence, it is necessary to remove the continuum signal in order to obtain pure gaseous emissions. A Sun spectrum or a solar analogue spectrum is used to remove the continuum signal. In this work, a solar analogue star, HD 19445, observed using the same instrument under the same settings, was used. The observed solar analogue spectrum is initially normalised and scaled to the comet continuum flux. A ratio of the polynomial fit to both spectra (using the continuum windows mentioned in \cite{ivanova_2021}) corrects for the redder dust of the comet. Multiplying the scaled solar spectrum by the polynomial ratio gives the comet continuum spectrum, which is then subtracted to extract the pure emission spectrum. The corresponding pure emission spectra extracted from the photocentre and the extreme points of the slit in the anti-tailward and tailward directions marked with the prominent emissions detected (for the centre) and regions of dissimilarity (for extreme regions) are shown in Figure \ref{F3_centre_ions}.

Furthermore, to facilitate the computation of the relative abundance of the detected ionic emissions along the spatial axis, multiple equidistant apertures (1, 2, 3, 4, and 5) of 12\arcsec, as shown in Figure \ref{F3_aperture}, were used. The sunward direction has been marked to clarify that the multiple apertures were extracted in the tailward (anti-sunward) direction. 

The multiple aperture extraction along the spatial axis in the anti-sunward direction shows that the ionic species corresponding to different molecular species start to be visible at about 3$\sigma$ level ($\sigma$ being the error in the continuum) for aperture distance beyond $\sim$35 000 km from the photocentre. The analysis of the column density profile of the major emissions (using Equation \ref{cd}), shows that the column densities of C$_3$ and NH$_2$, dropped steeply to become negligible beyond a distance of $\sim$20 000 km from the photocentre (see Figure \ref{F3_CD}). This sharp decline, along with the coincidental orientation of the strong ion tail with the slit direction, facilitated the detection of these species far from the photocentre (see Figure \ref{F3_tail}).

\subsubsection{Haser factor computation}\label{HF}
With long-slit spectroscopic observations of comets, only a portion of the total coma is observed. Therefore, the observed flux density of each molecular species must be extrapolated to determine the total flux density for the entire coma. Using the Haser coma outflow model, \cite{Fink1996} provides a factor known as the Haser factor, which is the ratio of the total number of molecules within the observed aperture to the total number of molecules in the entire coma. The reciprocal of this factor, the Haser correction, can be used to extrapolate the observed flux density and estimate the molecular abundance in the whole coma. The Haser model assumes a spherically symmetric coma with a uniform outflow of gas. A Web calculator provided by Dr. Schleicher on his website\footnote{\url{https://asteroid.lowell.edu/comet/}} can compute this factor for a circular aperture centred on the nucleus for a set of molecules (CN, C$_2$, C$_3$, NH, and OH).
     \begin{figure}[h!]

\centering
\includegraphics[width=0.8\linewidth]{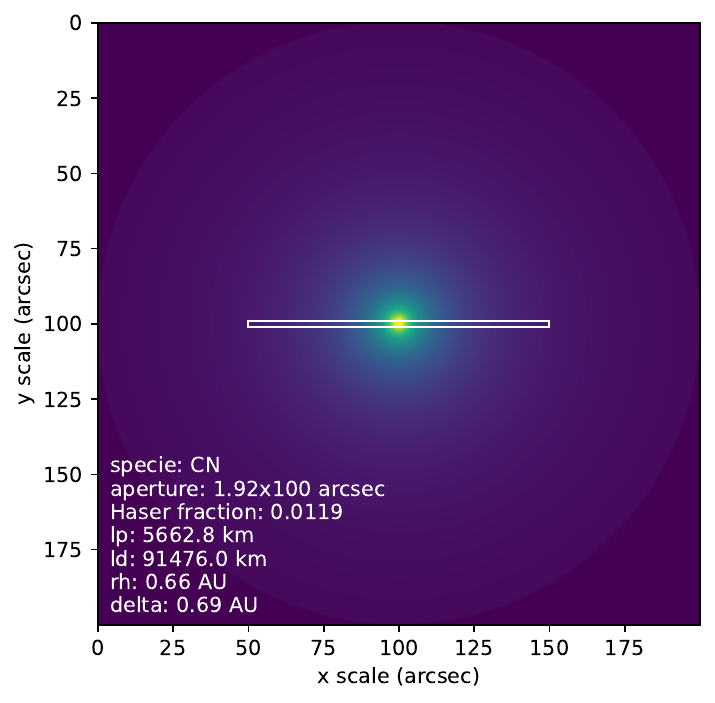}
\caption{Illustration of the rectangular slit for which the Haser factor computation is performed from a 2D image of the column density from the Haser model.}
         \label{haser_image}
   \end{figure}
   
However, in this work, we are using a rectangular slit, which makes it less straightforward to compute the Haser factor for an equivalent circular aperture. In addition, we are computing the production rates of NH$_2$ and H$_2$O for which the Haser factors are unavailable from Schleicher's website. Hence, we have developed a Python code that uses the activity module available within the sbpy package \citep{sbpy} to create a 2D image of the column density from the Haser model and use the photutils package \citep{photutils} to extract the Haser fraction for any given aperture. We cross-checked our results with the web calculator mentioned above for circular apertures, and they match decimal places with a better approximation.
   
This code can compute Haser factors for rectangular apertures of user-defined sizes with the comet at the centre and off-centred apertures. Figure \ref{haser_image} illustrates a similar scenario in which the Haser factor calculation of the CN molecule is carried out for an aperture of size of 100\arcsec~in a slit of width 1.92\arcsec. In this manner, the Haser factor can be computed for any molecule with known scale lengths for a user-defined aperture at a particular geocentric and heliocentric distance.
   
\section{Results}\label{discussion}
In this section, we present the light curve in different filters and the evolution of the comet activity and composition after its perihelion passage from photometric observation. We also discuss the results obtained from the spectroscopic observations.

\subsection{Photometry}\label{trappist_result}
We began the observation of 20F3 three weeks after perihelion on July 22, 2020 (\rh=0.63 au), as it was getting higher in the northern sky, at least 10 degrees above the horizon. We followed its activity with TN until September 10, 2020 (\rh=1.61 au). All species were detected since the first day of observation. As the comet was very bright around perihelion, short exposures were used to avoid saturating the images. The coma did not show any sign of outburst or increase in elongation, which is often associated with disruption of the nucleus at small heliocentric distances.

\subsubsection{Production rates and dust activity}
\begin{figure}[h!]
	\centering	\includegraphics[width=0.85\linewidth]{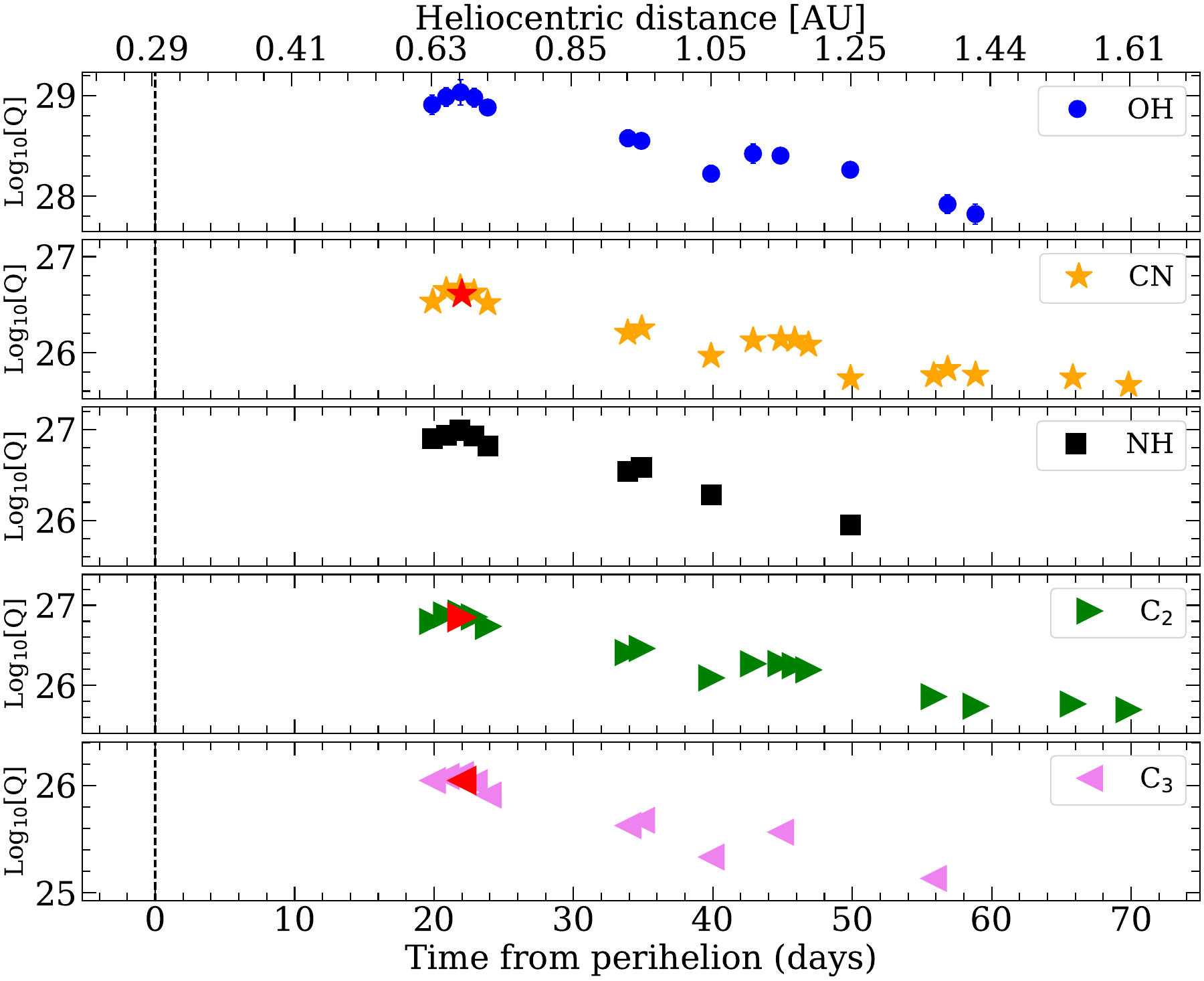}
	\caption{Logarithmic production rates of OH, NH, CN, C$_3$, and C$_2$,   detected in 20F3, as a function of time and heliocentric distance. The vertical dashed line indicates the perihelion on July 3, 2020, at 0.29 au. The red mark in each symbol represents the corresponding measurements from the spectroscopic observation.} 
	\label{fig:Qs20F3}
\end{figure}

\begin{figure}[h!]
	\centering	\includegraphics[width=0.8\linewidth]{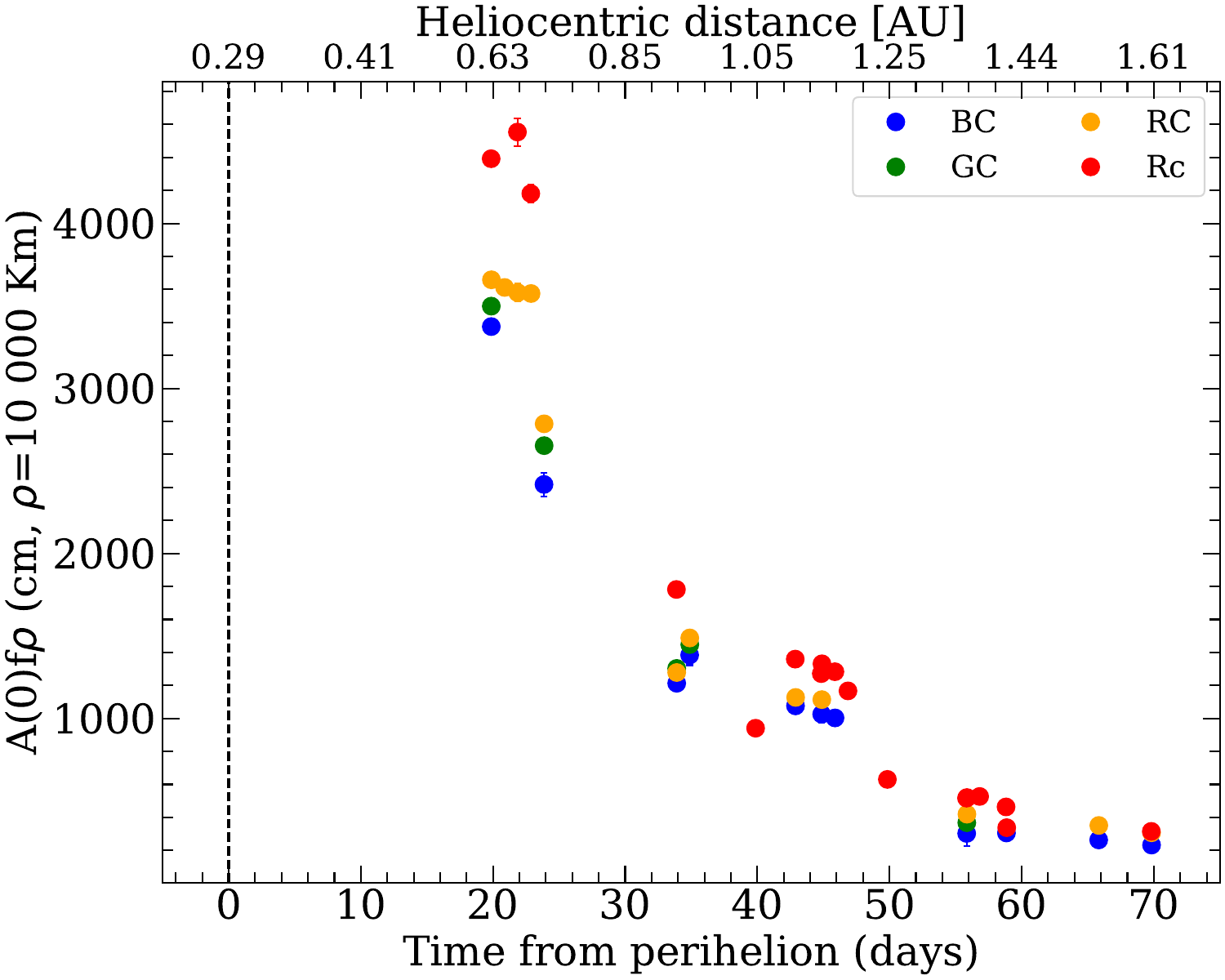}
	\caption{The A(0)f$\rho$ parameter of 20F3 at 10.000 km from the broad- and narrow-band filters as a function of time and heliocentric distance.} 
	\label{fig:afrho20F3}
\end{figure}

The derived production rates for each detected gas species, the A(0)f$\rho$ dust parameter and their absolute errors are given in Table \ref{tab:20F3}. Their evolution as a function of time and heliocentric distance is shown in Figures \ref{fig:Qs20F3} and \ref{fig:afrho20F3}.

The maximum activity of the comet was measured on the third night of our observations after perihelion (July 24, \rh = 0.67 au), with an OH production rate of (1.08$\pm$0.31)$\times$10$^{29}$ molec/s, and then decreased rapidly as the comet went away from the Sun. Schleicher (private communication) derived an OH production rate of 1.02$\times$10$^{29}$ molec/s (Jul 28, \rh = 0.74 au) from the Lowell Observatory. This result is higher than our measurement of (7.63$\pm$1.12)$\times$10$^{28}$ molec/s from two days earlier (July 26, \rh = 0.71 au). We obtained an OH production rate of (1.80$\pm$0.34)$\times$10$^{28}$ molec/s on the 11 of August (\rh = 1.05 au), and Schleicher obtained $4.16\times10^{28}$. The aperture around the nucleus was 25,700 km for Schleicher, and we use an annular aperture across 10,000 km for our calculations. The difference in production rates is the result of the different apertures and methodology used for the computation. \cite{F3_heterogenous} obtained high-resolution infrared spectra of comet 20F3 over July 2020. They were able to detect many parent species such as: CO, OCS, HCN, C$_2$H$_2$, NH$_3$, NH$_2$, H$_2$CO, CH$_4$, C$_2$H$_6$, and CH$_3$OH. Future work can compare daughter and parent species such as CN and HCN, OH and H$_2$O, C$_2$ and C$_2$H$_2$ or C$_2$H$_6$ and see if they agree well, which would then strengthen their links.
\begin{figure}[ht!]
	\centering	\includegraphics[width=0.85\linewidth]{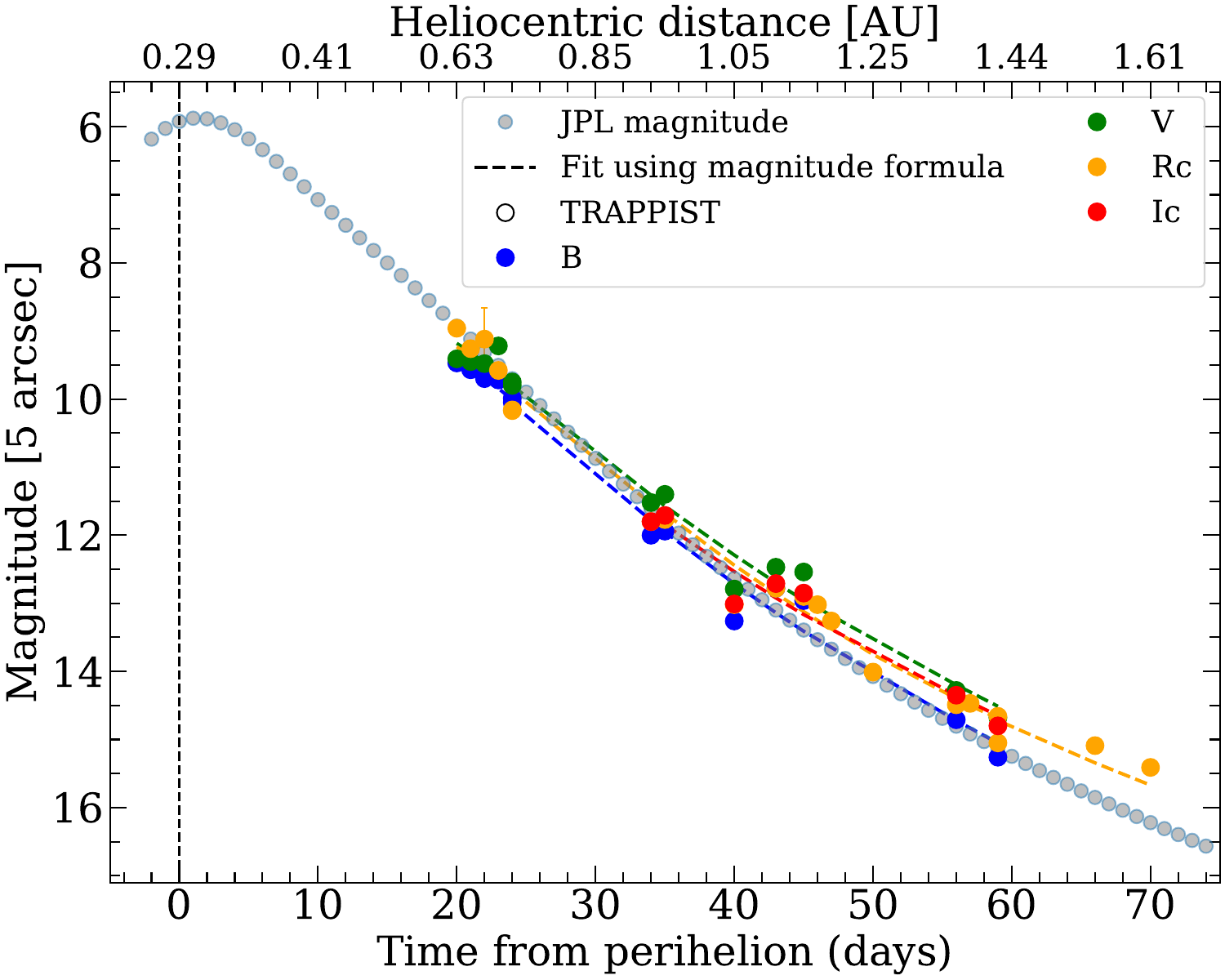}
	\caption{Optical light curve of 20F3 measured within a 5-arcsecond aperture as a function of time and distance to perihelion. The vertical dashed line indicates the perihelion at 0.29 au on July 03, 2020.} 
	\label{mag20F3}
\end{figure}

\begin{table*}[h!]
        \caption{Gas production rates and A(0)f$\rho$ parameter of 20F3 (NEOWISE).}
        \centering
        \label{tab:20F3}
        \resizebox{1\textwidth}{!}{%
            \setlength{\tabcolsep}{4pt}
            \renewcommand{\arraystretch}{1.17}
            \begin{tabular}{lcclllllllll}
                \hline
                \hline
                UT Date & $r_h$ & $\Delta$ & \multicolumn{5}{c}{Production rates (molec/s)} & \multicolumn{4}{c}{A(0)f$\rho$ (cm)} \\
                 & (au) & (au) & OH [$\times10^{28}$] & NH [$\times10^{26}$] & CN [$\times10^{26}$] & C$_2$ [$\times10^{26}$] & C$_3$ [$\times10^{26}$] & BC & RC & GC & Rc \\
                \hline
2020-07-22 & 0.63 & 0.69 & 8.13$\pm$1.79 & 7.91$\pm$0.73 & 3.40$\pm$0.25 & 6.28$\pm$0.16 & 1.11$\pm$0.07 & 3375$\pm$12 & 3658$\pm$3 & 3499$\pm$7 & 4392$\pm$4  \\
2020-07-23 & 0.65 & 0.69 & 9.71$\pm$2.04 & 8.63$\pm$0.92 & 4.47$\pm$0.29 & 7.58$\pm$0.17 & 1.21$\pm$0.07 & --& 3612$\pm$3 & --& -- \\
2020-07-24 & 0.67 & 0.70 & 10.80$\pm$3.10 & 9.84$\pm$0.74 & 4.76$\pm$0.29 & 7.99$\pm$0.20 & 1.26$\pm$0.07 & --& 3580$\pm$54 & --& 4554$\pm$84  \\
2020-07-25 & 0.69 & 0.70 & 9.56$\pm$2.08 & 8.46$\pm$0.63 & 4.22$\pm$0.20 & 7.16$\pm$0.15 & 1.06$\pm$0.05 & --& 3575$\pm$42 & --& 4181$\pm$52  \\
2020-07-26 & 0.71 & 0.71 & 7.63$\pm$1.12 & 6.58$\pm$0.52 & 3.27$\pm$0.13 & 5.46$\pm$0.09 & 0.81$\pm$0.03 & 2418$\pm$71 & 2786$\pm$25 & 2653$\pm$45 & -- \\
2020-08-05 & 0.93 & 0.91 & 3.77$\pm$0.72 & 3.46$\pm$0.35 & 1.61$\pm$0.08 & 2.57$\pm$0.06 & 0.42$\pm$0.02 & 1213$\pm$10 & 1278$\pm$5 & 1303$\pm$7 & 1782$\pm$5  \\
2020-08-06 & 0.95 & 0.93 & 3.55$\pm$0.56 & 3.85$\pm$0.28 & 1.79$\pm$0.08 & 2.89$\pm$0.06 & 0.47$\pm$0.02 & 1384$\pm$61 & 1488$\pm$40 & 1448$\pm$53 & -- \\
2020-08-11 & 1.05 & 1.08 & 1.67$\pm$0.32 & 1.91$\pm$0.19 & 0.92$\pm$0.04 & 1.24$\pm$0.03 & 0.21$\pm$0.01 & --& --& --& 941$\pm$3  \\
2020-08-14 & 1.11 & 1.17 & 2.65$\pm$0.57 & --& 1.34$\pm$0.07 & 1.86$\pm$0.04 & --& 1076$\pm$18 & 1127$\pm$10 & --& 1359$\pm$9  \\
2020-08-16 & 1.15 & 1.23 & 2.53$\pm$0.42 & --& 1.38$\pm$0.06 & 1.87$\pm$0.04 & 0.37$\pm$0.01 & 1026$\pm$50 & 1114$\pm$33 & --& 1331$\pm$26  \\
2020-08-17 & 1.17 & 1.26 & --& --& 1.36$\pm$0.06 & 1.76$\pm$0.04 & --& 1003$\pm$33& --& --& 1283$\pm$15  \\
2020-08-18 & 1.19 & 1.29 & --& --& 1.21$\pm$0.05 & 1.56$\pm$0.03 & --& --& --& --& 1167$\pm$11  \\
2020-08-21 & 1.25 & 1.39 & 1.83$\pm$0.20 & 0.89$\pm$0.11 & 0.54$\pm$0.03 & --& --& --& --& --& 630$\pm$12  \\
2020-08-27 & 1.36 & 1.58 & --& --& 0.58$\pm$0.03 & 0.72$\pm$0.02 & 0.14$\pm$0.01 & 303$\pm$77 & 420$\pm$44 & 368$\pm$52 & 521$\pm$22  \\
2020-08-28 & 1.38 & 1.61 & 0.83$\pm$0.18 & --& 0.67$\pm$0.03 & --& --& --& --& --& 527$\pm$20  \\
2020-08-30 & 1.41 & 1.68 & 0.66$\pm$0.15 & --& 0.59$\pm$0.03 & 0.55$\pm$0.02 & --& 306$\pm$25 & --& --& 338$\pm$11  \\
2020-09-06 & 1.54 & 1.90 & --& --& 0.55$\pm$0.03 & 0.58$\pm$0.02 & --& 263$\pm$35 & 351$\pm$20 & --& -- \\
2020-09-10 & 1.61 & 2.02 & --& --& 0.46$\pm$0.03 & 0.49$\pm$0.03 & --& 232$\pm$6 & 307$\pm$4 & --& 315$\pm$4  \\

                \hline
                \hline
            \end{tabular}}
\tablefoot{The A(0)f$\rho$ values are computed at 10,000 km from the nucleus and corrected for the phase angle effect. The reported uncertainties correspond to 1$\sigma$ absolute errors.}
\end{table*}

\begin{figure*}[ht!]
\centering	
\includegraphics[width=0.81\linewidth]{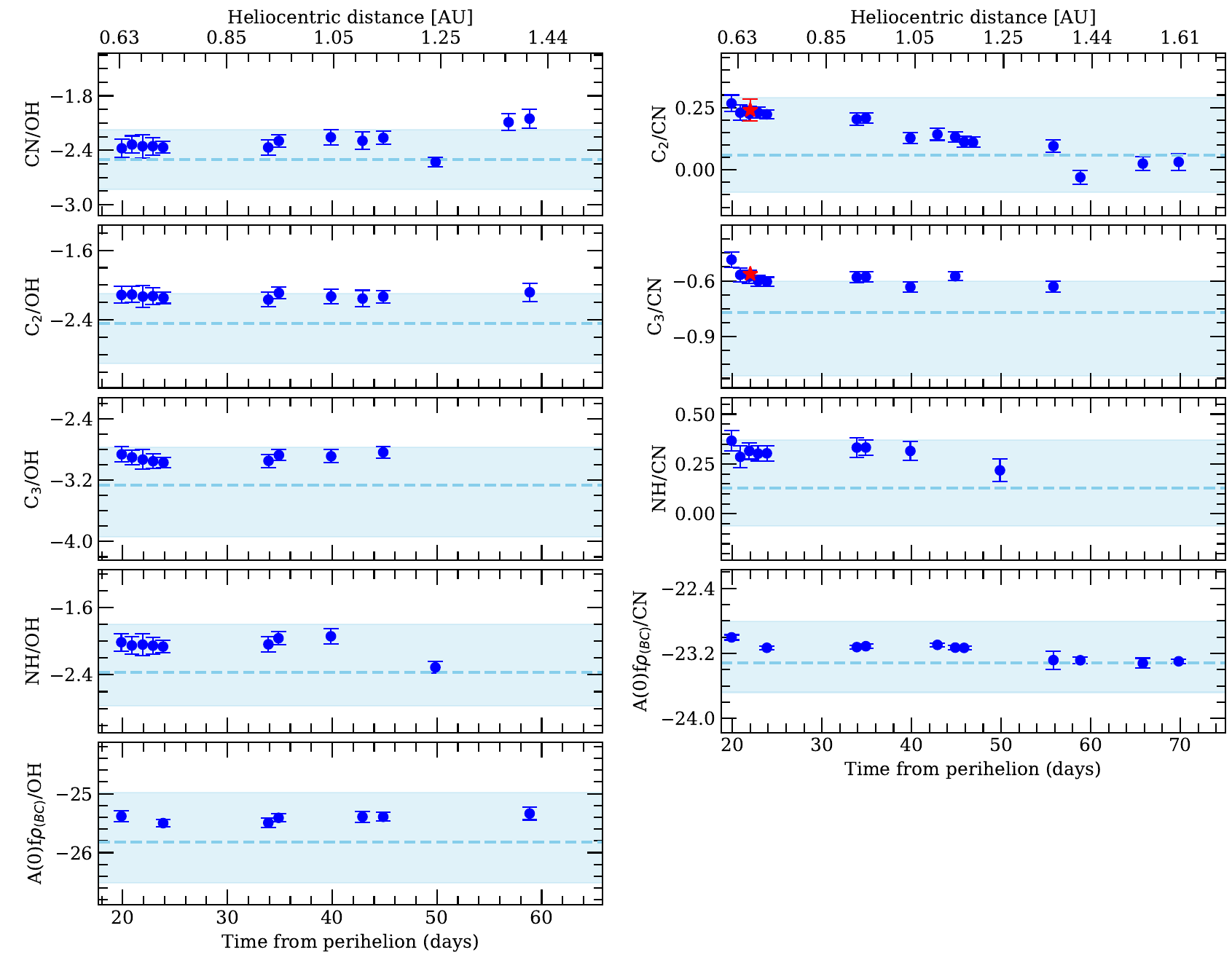}
	\caption{The logarithm of the production rates and dust-to-gas ratios of C/2020 F3 with respect to CN and to OH  as a function of time and heliocentric distance. The blue horizontal dashed line and the shaded region represent the mean ratios and range of ratios for comets with typical carbon composition as defined in \cite{Ahearn_85}. The red mark in a few panels represents the corresponding measurements from the spectroscopic observation. The errorbars correspond to 1$\sigma$ absolute errors.} 
    \label{prod_ratios}
\end{figure*}

Figure \ref{mag20F3} shows the evolution of the magnitude in different filters, the B, V, Rc, and Ic broadband filters, as a function of time, compared to the JPL ephemeris magnitude. The standard magnitude model "M = $M_{0}$ + 5 $\times$ log($r_{g}$) + 2.5 $\times$ n $\times$ log($r_{h}$) ", was used with the best-fit parameters; where n is the activity index, $M_{0}$ is the absolute magnitude and $r_{g}$ and $r_{h}$ are the geocentric and heliocentric distances, respectively. The best-fit parameters we obtained are: $M_{0(B)}$ = 12.32, $n_{(B)}$ = 4.22; $M_{0(V)}$ = 11.92, $n_{(V)}$ = 3.85; $M_{0(Rc)}$ = 12.06, $n_{(Rc)}$ = 4.03; and $M_{0(Ic)}$ = 12.18, $n_{(Ic)}$ = 3.53, for B, V, Rc, and Ic, respectively. The light curve shows no small-scale deviations, indicating that the activity evolution remained stable, with a noticeable fast drop after perihelion.

Figure \ref{prod_ratios} illustrates the relative abundances of the various daughter species and the dust/gas ratios with respect to CN and OH. Even though there are slight variations in the production rate ratios over the heliocentric distance range of 0.63-1.61 au, they remain well above or within the range defined by \cite{Ahearn_85} for comets with typical carbon composition. We obtained an average value of Log$_{10}$[Q(C$_2$)/Q(CN)]=0.14$\pm$0.02, which is in the high range of typical values for normal comets (0.06$\pm$0.10) as reported by \cite{Ahearn_85}. The mean value of dust/gas ratio (Log$_{10}$(A(0)f$\rho$/OH) = -25.42$\pm$0.40 and Log$_{10}$(A(0)f$\rho$/CN) = -23.17$\pm$0.11) is also well within the range observed for gas-rich comets. Hence, these ratios indicate that comet 20F3 has a typical coma composition with dust/gas ratios similar to typical gas-rich comets.

\subsubsection{Coma morphology and rotation period}
Coma features and jets in the CN narrowband images were investigated by applying azimedian and rotational \citep{larson_sekanina} filtering. The first CN series was obtained during the night starting on July 22. In azimedian-filtered images, two main jets separated by around 90$^\circ$, directed North and East on the plane of the sky, are distinguishable. A possible third jet oriented West can be observed in the rotational filtered image. A second series taken the night starting July 23 at the same time of the night shows similar features with similar orientation, indicating a rotation period close to an alias of 24h. Observations on the nights starting July 24, 25 and 26 show apparent disconnection between the Northern jet and the optocenter, as well as a radial outward translation of the jet feature, possibly indicating changes in the activity regions. Jets could be observed in our images until mid-August but cannot be directly linked to the July 22-23 jets, as new jets seem to be appearing and others fading out between observations.
\begin{figure}[h!]
\centering	\includegraphics[width=0.75\linewidth]{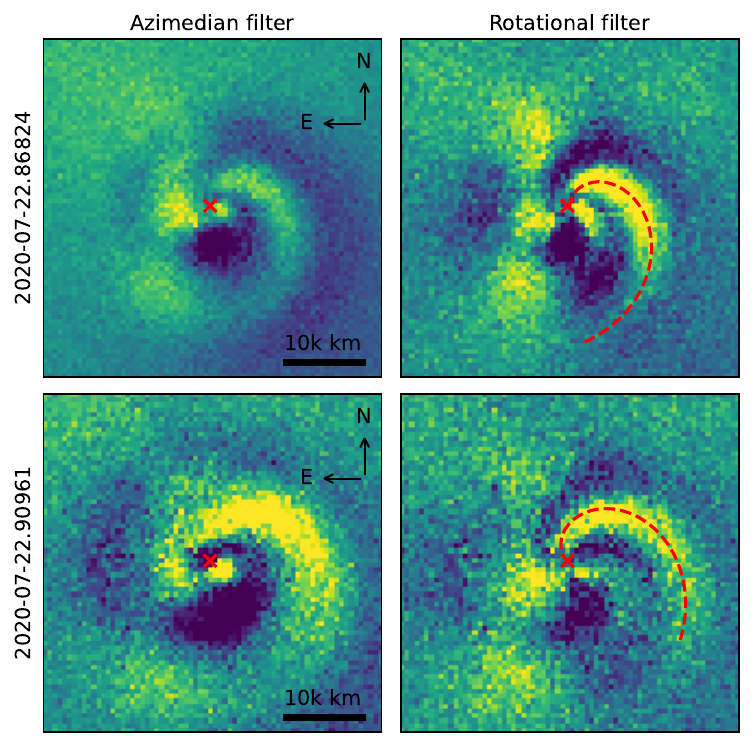}
	\caption{Azimedian (left) and rotational (right) CN narrow-band filtered images selected the night starting July 22, one hour appart. Images are oriented North up and East left. The red spirals follow the Northern jet that has been used for the rotational and velocity analysis.} 
    \label{rotation}
\end{figure}

To obtain the rotation period from the jets, we first did a Lomb-Scargle periodogram over the different consecutive nights. The periodic analysis was done with the angle between the tangent of the most prominent jet (oriented North when our images were taken; see Figure \ref{rotation}), the optocenter and the (arbitrarily chosen) East on each of the images. Given the difficulty to relate given jets between different nights, the analysis was limited to the nights of July 22 and 23, totalling 9 images. Periodogram analysis shows that the best matching frequencies (above a false alarm positive of 0.01) are 22.6 h and the aliases: 7.5 h, 4.5 h, and 3.8 h. This degeneracy is expected because the observations were made around the same time each night and the sample only contained images from two nights. However, from the image series of the night of 22 July, we can observe a rotation of the spirals of ~50$^\circ$ over one hour (see Figure \ref{rotation}), which confirms the period of $\sim$7.2h).

To complement the periodic analysis, the variation of the jet angle was measured over the night in series from the night starting July 22 (totalling 6 images) and July 26 (totalling 12 images). This method yielded periods of 7.91 $\pm$ 0.77 hours and 6.65 $\pm$ 0.81 hours, respectively. The given values are the average of the measurements on individual images over the night, with the uncertainty as the standard deviation. The obtained periods are in agreement with the two nights and with the periodic analysis. Our measurements are also consistent with the period of 7.8 $\pm$ 0.2 h derived by \cite{manzini_rotation} from the modelling of the coma dust; and 7.28 $\pm$ 0.03 h derived from features in C$_2$ narrow band images by \cite{Drahus2020ATel}. 

The associated CN outflow velocity in the plane of the sky could be obtained by fitting an archimedian spiral to the jet, assuming a constant CN outflow velocity. An Archimedean spiral of the form $r = b\theta$, where $r$ and $\theta$ are polar coordinates and $b$ determines the spacing between spiral arms, is fitted to the coma. The quantity $2\pi b$ corresponds to the projected distance between successive arcs along an axis through the nucleus, representing the distance the gas travels during one nucleus rotation. Using the known rotation period, the outflow velocity can be estimated. We could derive CN outflow velocities of 2.39 $\pm$ 0.23 km/s (July 23) and 2.44 $\pm$ 0.30 km/s (July 26). The derived velocity is seen to be on the higher side compared to the usually assumed velocity of 1 km/s or that obtained using the relation v=0.85/$\sqrt{r_h}$ \citep{cochran_30years}. However, it seems that this relation may not be applicable to all comets. \cite{Tseng_velocity} reported expansion velocities up to 2 km/s at 1 au for C/1995 O1 (Hale-Bopp) based on 18 cm OH line profiles, with similarly high velocities observed in 1P/Halley and C/1996 B2 (Hyakutake). Following the velocity–distance power law derived for Hale-Bopp in \cite{Tseng_velocity} to 0.65 au yields a velocity of 2.68 km/s, which is similar to our results for 20F3.

\subsection{Spectroscopy}
As seen in the top panel of Figure \ref{F3_centre_ions}, the neutral molecular species (CN, CH, C$_2$, C$_3$ and NH$_2$) 
are dominant in the spectrum extracted from the photocentre. However, the spectrum extracted from the tailward region of the cometary coma exhibits differences in emissions in multiple areas of the spectrum. The absence of certain emissions in the anti-tailward directions confirms their cometary origin. Most of the new emissions detected in the tailward spectrum, marked in the lower panel of Figure \ref{F3_centre_ions}, were mainly ionic emissions of different molecular species. The following subsections briefly discuss the computation of production rates of the neutral molecular species, the identification of the detected ionic emissions, the computation of their column densities, the variation of their abundance ratio along the spatial axis, and their implications.

\subsubsection{Production rates of neutral molecular species}
\begin{figure}[h!]
  \centering
   \includegraphics[width=0.90\linewidth]{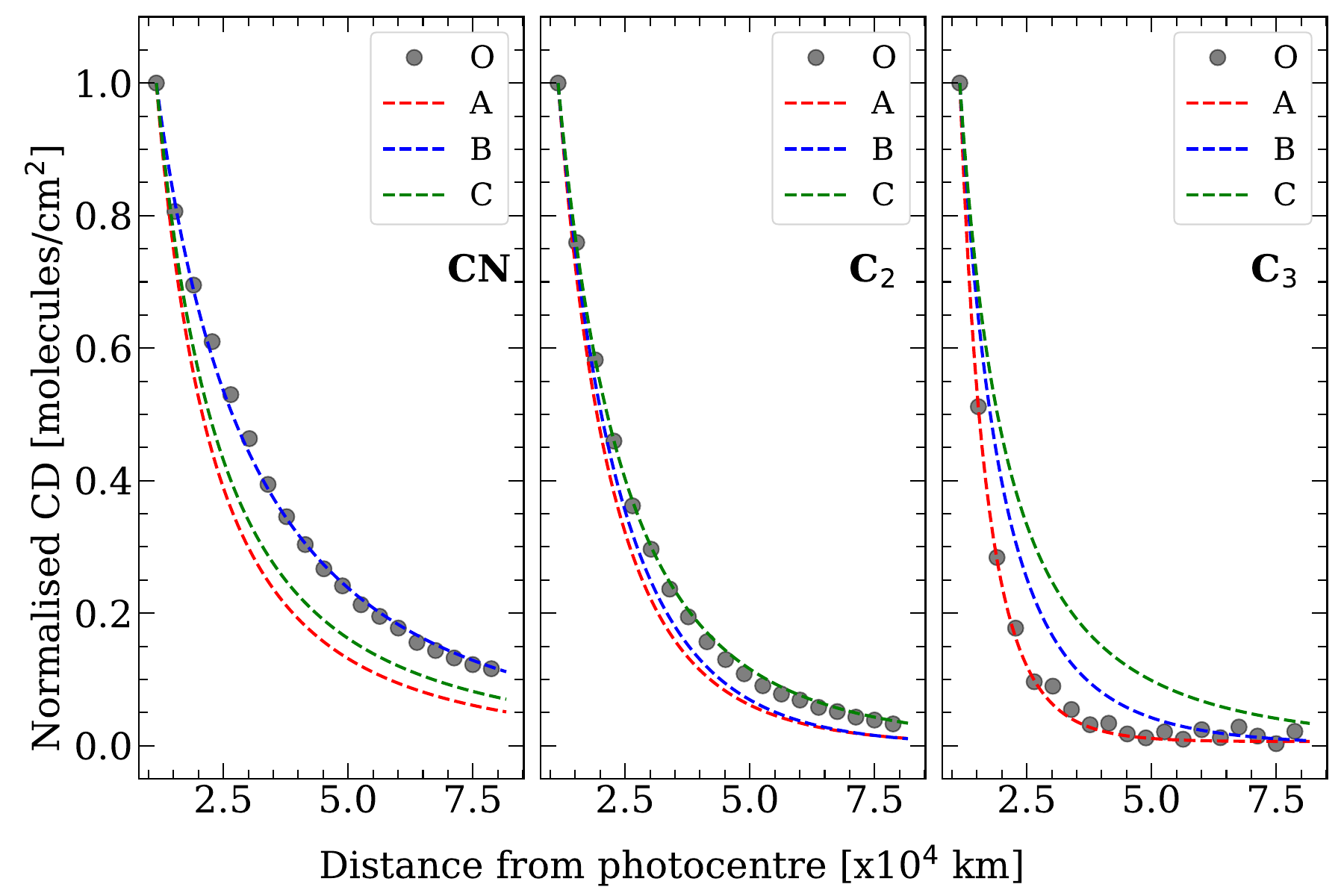}
      \caption{The comparison of Haser model fittings for the observed (O) column densities of CN, C$_2$ and C$_3$ making use of respective scale-lengths defined in \cite{Ahearn_85}(A), \cite{langland-shula}(B) and \cite{cochran_30years}(C).
              }
         \label{F3_CD_scalelengths}
   \end{figure}

In comet studies, production rates of species like CN, C$_2$, and C$_3$ usually come from fitting column density profiles with the Haser model (see \cite{156P_aravind} and \cite{langland-shula} for details) or from single aperture spectral extraction with Haser correction (see \cite{borisov_aravind} for details). However, scale lengths in active comets may vary, as already shown in the literature \citep[e.g.][and references there in]{Fink_scalelength, Cochran_scalelength, Fink_46P_scalelength}. Scale lengths from different studies (e.g. \cite{Ahearn_85}, \cite{cochran_30years}, \cite{langland-shula}) were chosen to illustrate this situation (see Figure~\ref{F3_CD_scalelengths}) and highlight the limitations of using generic scale lengths to represent the complex and changing chemistry of an active coma.

Developing a custom set of scale lengths for 20F3’s activity would improve our understanding of species and complex coma chemistry; however, this is beyond the scope of the present study, as our goal here is to derive production rates using the most commonly adopted scale lengths to enable comparison of 20F3 with other comets. It should be noted that, in the case of 20F3, although the fit quality of the Haser model varied with the choice of scale lengths, the resulting production rates were generally found to remain consistent within observational uncertainties.

In this work, to compute the production rates from spectroscopic data, single aperture spectral extractions were used along with the equation,
\begin{equation}\label{prod_rate}
    Q = \frac{4 \pi \Delta^2 v_{out}}{gl_d}\times Flux \times HC,
\end{equation}
where $\Delta$ is the geocentric distance in au, g is the fluorescence efficiency (ergs/molecules/s), v$_{out}$ is the outflow velocity assumed to be 1 km/s, and l$_d$ is the scale length of the daughter molecule. Here, Flux in Equation \ref{prod_rate} is the total spectral flux density of the molecule of interest within the defined bandpass (taken from \cite{langland-shula} and \cite{Fink_comet_survey_2009}), and HC is the Haser correction, defined as the inverse of the Haser factor. Haser factor, the ratio of the number of molecules of the emitting species within an arbitrary aperture to the total number of molecules if the aperture were extended to infinity, is computed as mentioned in Section \ref{HF}.
\begin{table}[h!]
    \caption{Scale lengths and fluorescence efficiencies adopted for the different molecules.}
    \label{parameters}
    \centering
    \setlength{\tabcolsep}{3.5pt}
        \resizebox{\linewidth}{!}{%
    \begin{tabular}{cccccc}
    \hline \hline
        Molecule & l$_p$ & l$_d$ & g & Ref.  \\
        & ($\times$10$^4$ km) & ($\times$10$^4$ km) & (ergs/mol/s) &\\
        \hline
         CN & 1.3 & 21 & 2.62$\times10^{-13}$ & 1,3  \\
         C$_2$& 2.2 & 6.6 & 4.5$\times10^{-13}$ & 1 \\
         C$_3$& 0.28 & 2.7 & 1.0$\times10^{-12}$ & 1  \\
         NH$_2$& 0.49 & 6.2 & 9.22$\times10^{-15}$ & 2 \\
         H$_2$O [OI]& 8.0 & -- & -- & 2\\
         \hline
         \hline
    \end{tabular}}
    \tablefoot{The values are for 1 au heliocentric distance.}
    \tablebib{
            (1)~\citet{Ahearn_85}; (2) \citet{Fink_comet_survey_2009}; (3) \cite{schleicher_CN_2010}
            }
\end{table}

Since the scale-lengths defined in \cite{Ahearn_85} are widely used for computing the production rates of comets, we also adopt the same values for molecules CN, C$_2$ and C$_3$  to maintain consistency and ensure uniformity in comparison between different data sets and observational methods. The values are then scaled to $r_h^{2}$, $r_h$ being the heliocentric distance. Although the fluorescence efficiencies of C$_2$ and C$_3$ have been taken from \cite{Ahearn_85} and scaled to $r_h^{-2}$, that of CN is obtained by performing a double interpolation in the table provided by \cite{schleicher_CN_2010}, which contains the g factor of CN for different heliocentric distances and velocities. The scale length and fluorescence efficiency of the NH$_2$ (0-10-0) band are taken from \cite{Fink_comet_survey_2009} and scaled to $r_h^{2}$ and $r_h^{-2}$ respectively (see Table \ref{parameters} for details).
\begin{figure}[h!]
  \centering
   \includegraphics[width=0.75\linewidth]{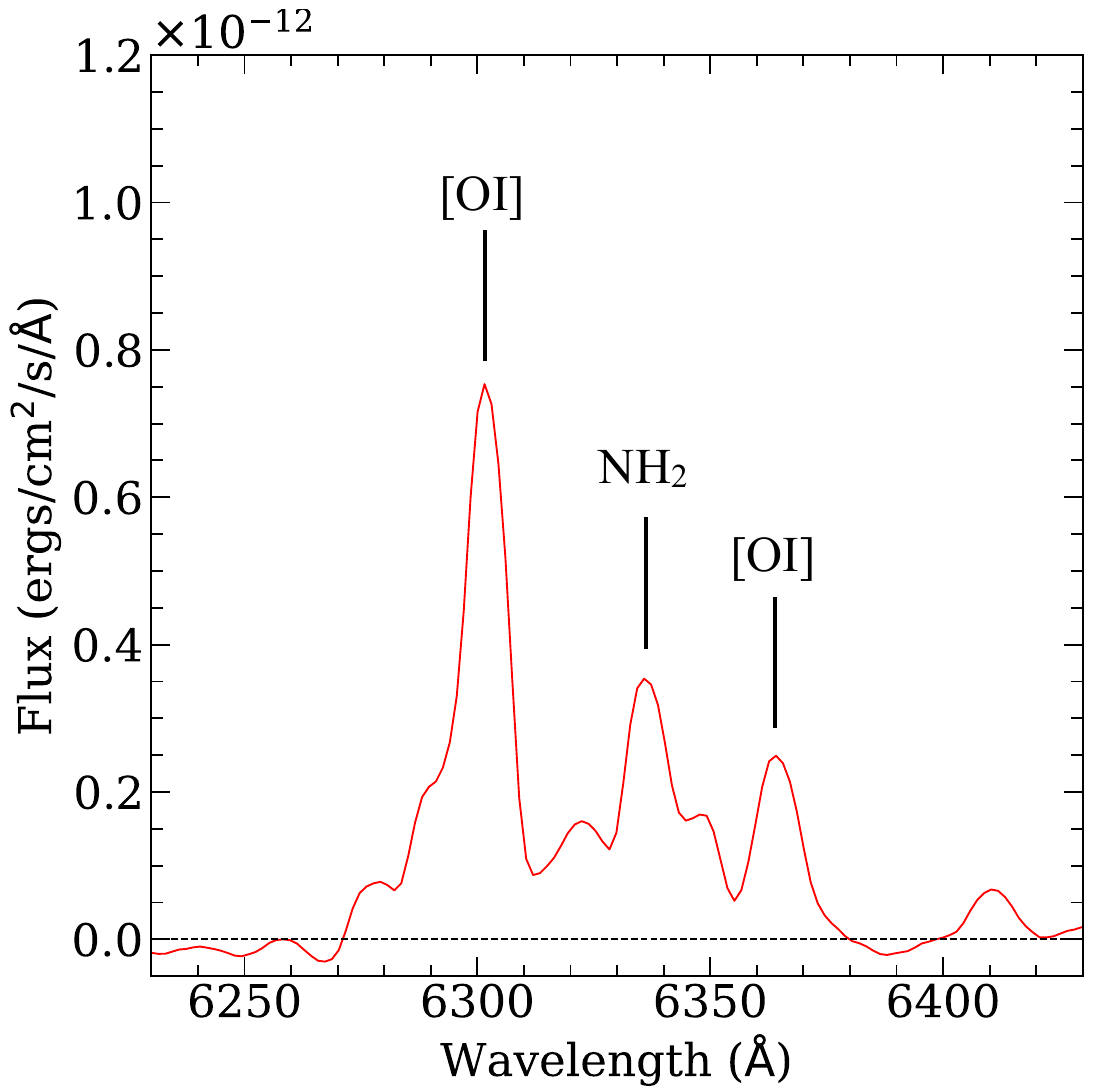}
      \caption{Illustration of the 6335\,\AA~NH$_2$ peak that is used to decontaminate the NH$_2$ contribution in the [OI] 6300\,\AA~line.
              }
         \label{F3_OI}
   \end{figure}   
   
We also attempt to compute the water production rate using the spectral flux density of the forbidden Oxygen line (hereafter [OI]) at 6300\,\AA. This approach has previously been used in many cases \citep[eg., ][]{schultz_1992_oxygen, Fink1996, Morgenthaler_2001,Morgenthaler_2007, Mckay_2012_oxygen, 67P_muse_oxygen, Mckay_borisov_oxygen}. In this work, since the observations are in low-resolution mode, there would be significant contamination from the sky and the nearby NH$_2$ bands. As mentioned in Section \ref{longslit},  the sky contamination has been effectively removed, leaving us with the NH$_2$ contamination. We follow the method described in \cite{Fink_comet_survey_2009} to remove the NH$_2$ contamination, where 0.9 times the spectral flux density of the (0-8-0) NH$_2$ band peak at 6335\,\AA~ is subtracted from that of the [OI] 6300\,\AA~line (see Figure \ref{F3_OI} for the mentioned peaks and Table \ref{parameters} for the adopted scale length).

The resulting production rate of Oxygen atoms in the ($^1$D) state, Q([OI]), is computed as 
\begin{equation}
    Q([OI]) = \frac{4}{3}(4\pi\Delta^2 F_{6300})\times HC,
\end{equation}

where $\Delta$ is the geocentric distance in cm, F$_{6300}$ is the decontaminated spectral flux density of the 6300\,\AA~[OI] line (in photons s$^{-1}$ cm$^{-2}$), and HC is the Haser correction to account for the emission not encompassed in the aperture \citep[e.g.,][]{Morgenthaler_2001, schultz_1992_oxygen}. Since [OI] emission is the result of a forbidden transition rather than fluorescence, if all these photons come from O($^1$D) created during the photodissociation of
H$_2$O and its daughter OH, the water production rate Q(H$_2$O) can be derived from O($^1$D) production rate, using the expression,
\begin{equation}
    Q(H_2O) =  \frac{Q([OI])}{BR1+(BR2*BR3)},
\end{equation}
where BR1, BR2, and BR3, chosen to be 0.05, 0.855, and 0.094, respectively (see \cite{67P_muse_oxygen} and references therein), are the branching ratios for the three reactions,\\

\noindent H$_2$O + h$\nu$ $\rightarrow$ H$_2$ + O($^1$D)\null\hfill(i)\\
H$_2$O + h$\nu$ $\rightarrow$ H + OH\null\hfill(ii)\\
OH + h$\nu$ $\rightarrow$ H + O($^1$D)\null\hfill(iii)\\

   \begin{table}[h!]
    \centering
    \caption{Production rates and rate ratios for the major neutral molecules in 20F3 observed on 2020-07-24. }
    \label{rates_ratios}
    \setlength{\tabcolsep}{4.5pt}
        \resizebox{\linewidth}{!}{%
    \begin{tabular}{cccc}
    \hline \hline
        Molecule & Q$(10^{26}$ molecules/s)  & log($\frac{Q}{H_2O}$)  & log($\frac{Q}{CN}$) \\
         \hline
         H$_2$O [OI] & 4930 $\pm$ 158  & 0 & 3.08 $\pm$ 0.01 \\
         CN & 4.06 $\pm$ 0.05 & -3.08 $\pm$ 0.01 & 0 \\
         C$_3$ & 1.11 $\pm$ 0.02 & -3.64 $\pm$ 0.02 & -0.56 $\pm$ 0.01 \\
         C$_2$ & 7.05 $\pm$ 0.7 & -2.84 $\pm$ 0.03 & 0.24 $\pm$ 0.02\\
         NH$_2$ & 5.88 $\pm$ 0.15  & -2.92 $\pm$ 0.02 & 0.16 $\pm$ 0.01\\
         \hline
    \end{tabular}}
\end{table}
Using the above-mentioned computation methods, production rates for CN, C$_2$, C$_3$, NH$_2$ and H$_2$O were computed for multiple aperture sizes varying from 90 to 120 arcseconds to minimise any possible effects of quenching within the inner coma (crucial in the case of H$_2$O) and non-uniform slit illumination towards ends of the slit. The average of the computed production rates and their standard deviation taken as the 1$\sigma$ absolute error, are listed in Table \ref{rates_ratios}.

\begin{figure}[h!]
  \centering
   \includegraphics[width=0.80\linewidth]{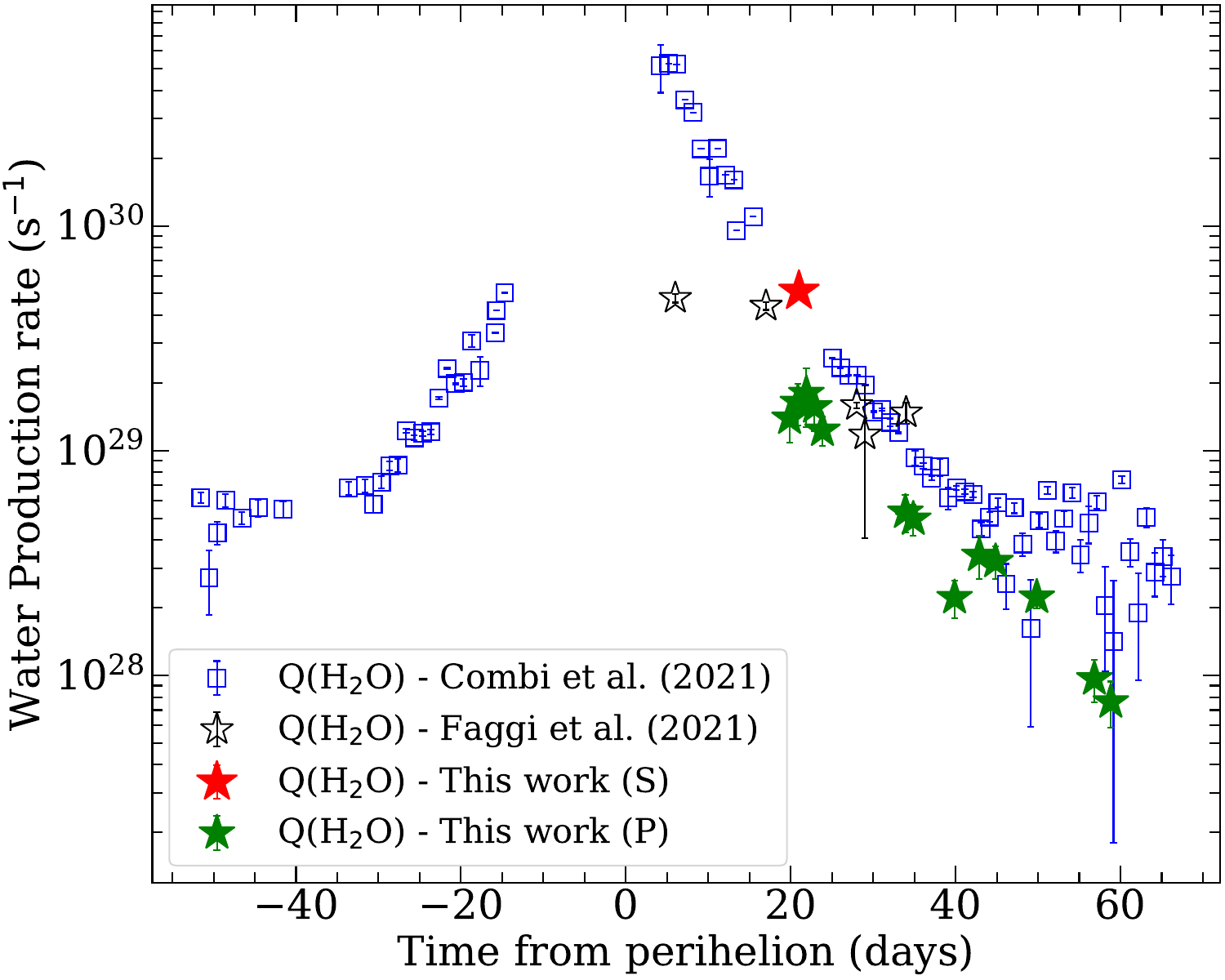}
      \caption{The comparison of water production of 20F3 computed in this work using the 6300\,\AA~[OI] line spectral flux density (S) and OH narrow band images (P) with those reported in \cite{C2020F3_combi_waterprod} and \cite{F3_heterogenous}. The errorbars correspond to 1$\sigma$ absolute errors.
              }
         \label{F3_h2o}
   \end{figure}

The computed production rates of CN, C$_2$, and C$_3$ are comparable to those derived from the TRAPPIST observations, as discussed earlier in Section \ref{trappist_result}. It has been observed that the water production rate computed in this work from the [OI] line, $\sim$20 days after perihelion, is in very good agreement with that reported by \cite{C2020F3_combi_waterprod} from SOHO/SWAN H$_{Ly-\alpha}$ observations and \cite{F3_heterogenous} from iSHELL/IRTF observations of H$_2$O lines in the Near-InfraRed (NIR) region (see Figure \ref{F3_h2o}). The water production rate calculated using the empirical conversion formula, Q(H$_2$O) = 1.36$\times$r$_h^{-0.5}\times$Q(OH) \citep{Schleicher1998}, is slightly on the lower side compared to the other results. A similar discrepancy in water production rates between TRAPPIST and SOHO values has previously been highlighted by \cite{opitom_lemmon}, \cite{dellorusso_OH_H2O}, and \cite{moulane_46p}. These inconsistencies between the water production rates reported from spectroscopic and narrow band photometric observations could be due to a possible residual NH$_2$ contamination within the [OI] 63000\,\AA~line, differences in adopted scale lengths, and the complexities of coma chemistry.

From this work, it can be inferred that the spectral flux density of the [OI] line can be used to calculate a robust upper limit of water production rates. Furthermore, the production rate ratios with respect to CN, mentioned in Table \ref{rates_ratios}, classify comet 20F3 as typical in composition according to the definition mentioned in \cite{Ahearn_85}. 
   
      \begin{figure*}[!h]
  \centering
   \includegraphics[width=0.75\linewidth]{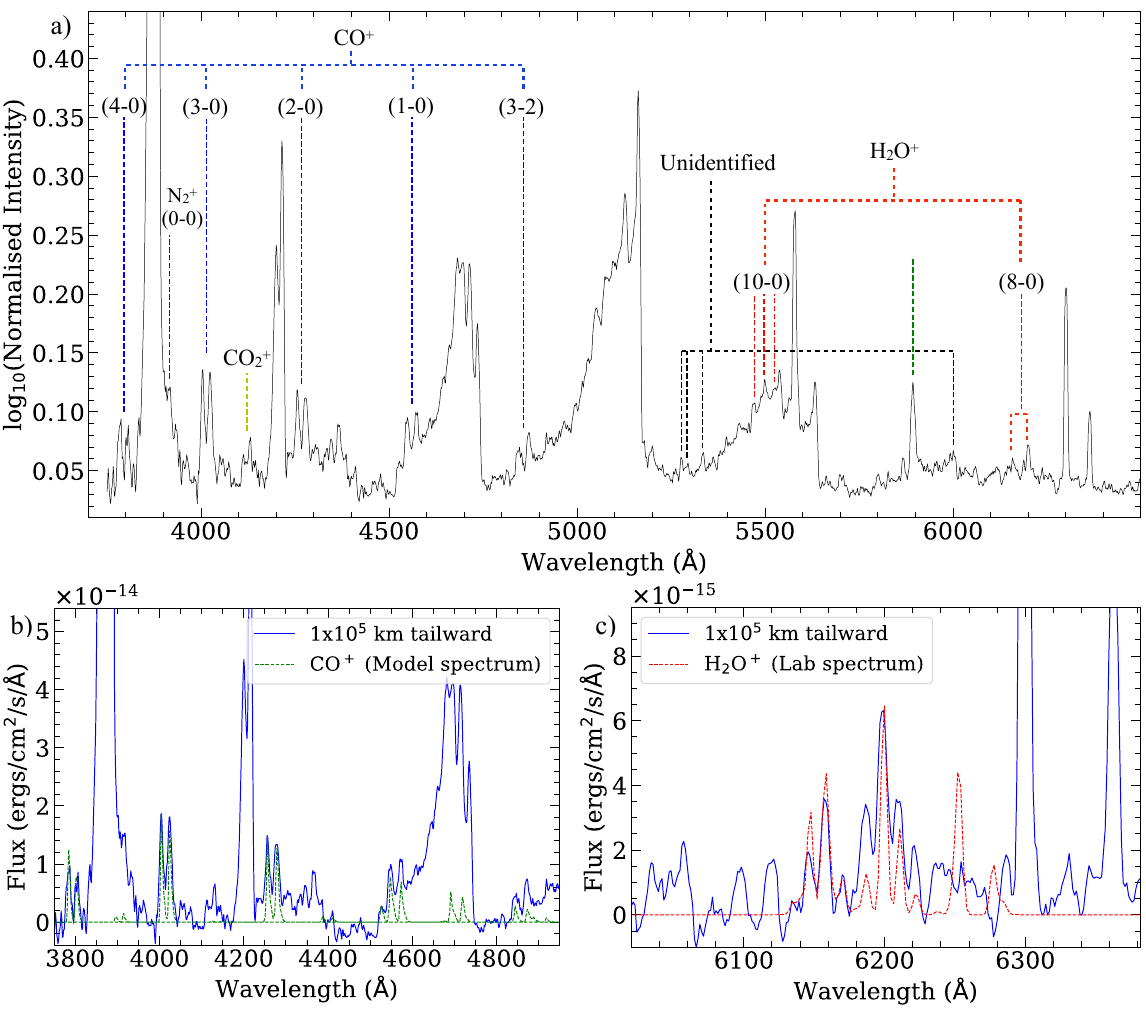}
      \caption{Spectrum of 20F3 extracted from the tailward direction, 1$\times10^5$ km away from the photocentre (a) depicting the detection of ionic emissions and few unidentified emissions; (b) overplotted with the modeled spectrum of CO$^+$ \citep{CO+_Rousselot} convolved with the instrument profile and (c) overplotted with the laboratory spectra of H$_2$O$^+$ \citep{Lew_h2op} convolved with the instrument profile.}
         \label{F3_ions}
   \end{figure*}
    
\subsubsection{Detected ionic emissions and Sodium}
Emissions from ionic species N$_2^+$, CO$^+$, CO$_2^+$, H$_2$O$^+$, as well as sodium (NaI) were detected in 20F3 while exploring the spatial axis in the tailward direction. The respective emissions and their corresponding bands are shown in panel (a) of Figure \ref{F3_ions}. The identified sodium emission belonging to the sodium doublet at 5889.95\,\AA~ (D2) and 5895.92\,\AA~ (D1), originating from the comet and the atmosphere of the Earth, are indistinguishable at our low spectral resolutions. With the sky contribution for an exposure time of three minutes being minimal for bright comets like C/2020 F3 and having performed proper sky removal, as explained earlier, the remaining observed NaI emission can be considered to be of cometary origin. The bottom panel of Figure \ref{F3_centre_ions} shows the excess sodium emission in the tailward direction that confirms its cometary origin. The ionic emissions were identified with the help of similar work reported in other publications \citep[eg.,][]{wyckoff_ionsincometails, Cochran_ions_N2+_CO2+, wyckoff_Halley_ions, wyckoff_unid, Kumar_R2, Hardy_atlas}. The steep drop in column densities of the major neutral species C$_3$ and NH$_2$ in the blue and red range (see Figure \ref{F3_CD})
facilitated the detection of the ionic species CO$^+$ and H$_2$O$^+$ respectively, far from the nucleus (see Figure \ref{F3_tail}).

The confirmation of emissions from CO$^+$ and H$_2$O$^+$ were obtained with the help of the model spectra of CO$^+$ \citep{CO+_Rousselot} and the laboratory spectra of H$_2$O$^+$ \citep{Lew_h2op} convolved with the instrument profile obtained from the FeAr calibration lamp lines (see panel (b) and (c) of Figure \ref{F3_ions}). It is presumed that excess emission in the tailward spectrum around 5500\,\AA~(see the bottom panel of Figure \ref{F3_centre_ions}) is due to the presence of the H$_2$O$^+$ (10-0) emissions at 5469.97\,\AA, 5479.78\,\AA~and 5488.78\,\AA, which has been identified with the help of the markings provided in \cite{wyckoff_ionsincometails} and the detailed line list provided in \cite{wyckoff_unid}.

\subsubsection{Ion column densities and abundance ratios}\label{ion_ratios}
The column density (N) of an ion, whose electronic transitions are excited by resonance fluorescence, is given by
\begin{equation}\label{cd}
    N = \frac{4\pi F}{g\Omega},
\end{equation}
where F is the integrated band flux density of the cometary ion of interest, g is the fluorescence efficiency, and $\Omega$ is the solid angle subtended by the aperture used at the comet. Fluorescence efficiencies (ergs/ion/s) at 1 au of 1.7$\times$10$^{-14}$ and 1.36$\times$10$^{-14}$ for the CO$^+$(3-0) and CO$^+$(2-0) bands respectively \citep{CO+_Rousselot}, 2.48$\times$10$^{-13}$ for the N$_2^+$(0-0) band \citep{N2plus_rousselot} and 2.1$\times$10$^{-14}$ for the H$_2$O$^+$(0-8-0) band \citep{N2plus_H2Op_g_lutz} has been adopted. In all cases, the fluorescence efficiencies were scaled to the appropriate heliocentric distance (\rh), following $g(r)$ = $g(1~au)/r_h^2$. From panel (a) of Figure \ref{F3_ions}, it is clearly seen that the N$_2^+$ band is highly blended with the CN emission, which makes it difficult to extract the actual flux. Special care has to be taken to extract only the flux density of the visible emission band head to compute the abundance ratio. Dedicated fluorescence modelling techniques are to be incorporated to estimate the underlying N$_2^+$ within the CN emission band. Table \ref{ion_flux} provides the flux densities and their absolute errors  for the detected ionic species for a 12\arcsec aperture at various distances from the photocentre.

\begin{table}[h]
    \renewcommand{\arraystretch}{1.5}
    \caption{Observed ion flux densities (ergs/cm$^2$/s).}
    \label{ion_flux}
    \centering
    \setlength{\tabcolsep}{4.5pt}
        \resizebox{\linewidth}{!}{%
  \begin{tabular}{lllll}
    \hline
    \hline
    Distance [km] & N$_2^+$ & CO$^+$(3-0) & CO$^+$(2-0) & H$_2$O$^+$(0-8-0) \\ \hline
    35809   & (1.51$\pm$0.17)$\times10^{-13}$  & (3.91$\pm$0.27)$\times10^{-13}$  & (1.49$\pm$0.30)$\times10^{-13}$  & (2.81$\pm$0.14)$\times10^{-13}$  \\ 
    53783   & (1.06$\pm$0.07)$\times10^{-13}$  & (2.48$\pm$0.12)$\times10^{-13}$  & (1.12$\pm$0.13)$\times10^{-13}$   & (1.41$\pm$0.22)$\times10^{-13}$  \\ 
    71688  & (8.99$\pm$1.16)$\times10^{-14}$ & (1.74$\pm$0.19)$\times10^{-13}$  & (7.71$\pm$1.15)$\times10^{-14}$  & (7.91$\pm$0.41)$\times10^{-14}$  \\ 
    89294  & (8.02$\pm$0.97)$\times10^{-14}$ & (1.57$\pm$0.16)$\times10^{-13}$  & (7.48$\pm$1.75)$\times10^{-14}$  &  (4.76$\pm$0.52)$\times10^{-14}$ \\ 
    103548  & (6.71$\pm$0.79)$\times10^{-14}$ & (1.48$\pm$0.13)$\times10^{-13}$  & (6.99$\pm$1.42)$\times10^{-14}$  & (3.69$\pm$0.84)$\times10^{-14}$  \\ \hline
  \end{tabular}}
  \tablefoot{The reported flux are for 12\arcsec~aperture at increasing distances from the photocentre.}
\end{table}

The value of N$_2$/CO for protosolar nebula was calculated to be 0.145$\pm$0.048 \citep{solar_nebula_N2_CO, lodders_solar_nebula, rubin_solar_nebula}. Since N$_2$ is the least reactive among nitrogen-bearing species and serves as the primary carrier of nitrogen in the solar nebula, a low abundance of elemental nitrogen in a comet must primarily result from the depletion of N$_2$ itself \citep{N2_feldman}. \cite{N2_CO_orbital} assert that the measured N$_2$/CO ratio is unrelated to the comet's dynamical history (number of passages into the inner Solar system) and therefore directly represents the volatile composition of the comet. \cite{owen_N2_CO} reported that the N$_2$/CO of comets that formed around a temperature of 50 K would be 0.06 if N$_2$/CO is 1 in the solar nebula. \cite{owen_N2_CO} also states that comets forming at higher temperatures, close to the vicinity of Jupiter, would have very less abundance of pristine ices such as N$_2$, CO, etc. This implies that the abundance of N$_2$ in a comet, measured through N$_2$/CO, is a direct evidence of the approximate temperature (and hence the location in the protosolar nebula) at which the comet was formed. Since N$_2$ cannot be detected directly in comets through spectroscopic observations, the N$_2^+$/CO$^+$ ratio is measured instead. This ratio is effectively equivalent to N$_2$/CO, assuming that N$_2$ and CO have similar ionisation efficiencies and are fully photodissociated in the coma. 
\begin{table}[h]
  \centering
    \renewcommand{\arraystretch}{1.5}
    \caption{Observed ion abundance ratios.}
    \label{ion_ratios}
    \setlength{\tabcolsep}{4.5pt}
        \resizebox{\linewidth}{!}{%
  \begin{tabular}{llll}
    \hline
    \hline
    Distance [km] & N$_2^+$/CO$^+$(3-0) & N$_2^+$/CO$^+$(2-0) & N$_2^+$/H$_2$O$^+$(0-8-0) \\ \hline
    35809   & 0.026$\pm$0.005  & 0.031$\pm$0.006  & 0.045$\pm$0.006 \\ 
    53783   & 0.029$\pm$0.003  & 0.028$\pm$0.004  & 0.063$\pm$0.008  \\ 
    71688  & 0.035$\pm$0.008  & 0.035$\pm$0.008  & 0.098$\pm$0.017  \\
    89294  & 0.035$\pm$0.007  & 0.032$\pm$0.009  & 0.142$\pm$0.033  \\
    103548  & 0.031$\pm$0.006  & 0.029$\pm$0.007  & 0.154$\pm$0.053  \\ \hline
  \end{tabular}}
  \tablefoot{The reported abundance are for 12\arcsec~aperture at increasing distances from the photocentre.}
\end{table}

Hence, taking into account the importance of this ratio, we used
\begin{equation}\label{ion_ratio}
    \frac{N_2^+}{CO^+}=\frac{g_{CO^+}}{g_{N_2^+}} \frac{F_{N_2^+}}{F_{CO^+}},
\end{equation}
to derive the average $N_2^+/CO^+$ = 0.03$\pm$0.01 for both the (3-0) and (2-0) bands of CO$^+$ (see Table \ref{ion_ratios} for the ratios computed for a 12\arcsec aperture at various distances from the photocentre). The variation of the abundance in the spatial direction along the slit was checked for these bands, and it was found to be consistent within the error bar for nucleocentric distances varying from $\sim$4$\times$10$^4$ km to $\sim$1$\times$10$^5$ km.
However, it was observed that the abundance ratio, $N_2^+/H_2O^+$, measured for the same distances, was seen to vary from $\sim$0.04 to $\sim$0.15, which is comparable to the values reported by \cite{N2plus_H2Op_g_lutz}, but lower by a factor of 10 for those reported by \cite{N2_H2O_modelling} based on photochemical models. As mentioned before, since the N$_2^+$ flux density is not measured for its entire wavelength range, we incorporate the fluorescence modelling technique to obtain a better approximation of the ratio. The details have been discussed in Section \ref{N2_model}.

 \subsubsection{Detection of CO$_2^+$ and unidentified emissions}\label{co2}
 Analysis of the tailward spectrum facilitated the identification of a few previously reported unidentified emissions. The unidentified emissions reported by \cite{wyckoff_unid} for comet Hyakutake (C/1996 B2) at locations 5273\,\AA, 5289\,\AA, 5333\,\AA~and 6003\,\AA~ were also observed in 20F3 (see panel (a) of Figure \ref{F3_ions}). Due to the availability of the very high-resolution emission atlas for 20F3 \citep{F3_highres} and the comet line atlas provided by \cite{Hardy_atlas}, the previously unidentified emissions observed at 5273\,\AA, 5333\,\AA, and 6003\,\AA~ are identified as possibly belonging to the C$_2$ bands, while the emission at 5289\,\AA~possibly belongs to NH$_2$. Subsequently, \cite{bodewits_H2O+} demonstrated that many of the previously unidentified emissions described in \cite{wyckoff_unid} can be attributed to transitions originating from the higher vibrational levels of H$_2$O$^+$ (A$^2$A$_1$ - X $^2$B$_1$). 

 \begin{figure}[h!] 
  \centering   \includegraphics[width=0.87\linewidth]{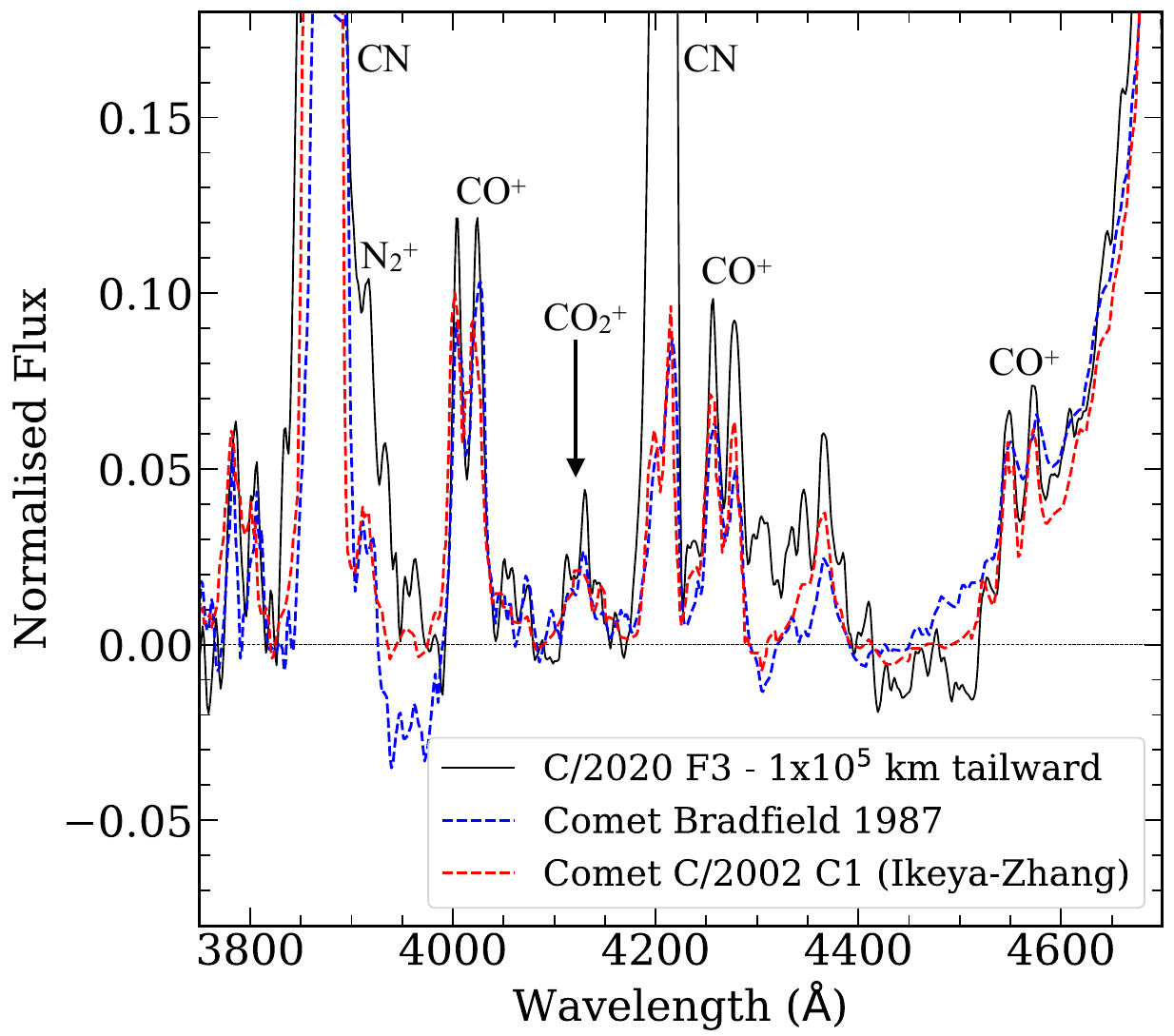}
      \caption{Tailward spectrum of 20F3 overlayed with those of comet Bradfield 1987 \citep{N2plus_H2Op_g_lutz} and comet Ikeya-Zhang \citep{Cochran_ions_N2+_CO2+} illustrating the detection of CO$_2^+$ emission at 4130\,\AA.
              }
         \label{F3_CO2p}
   \end{figure}  
   
In addition, in the past, there have only been a small fraction of known comets with significant detection of CO$_2^+$ emissions from ground-based observations. Comet 1P/Halley was one major candidate to have multiple reports of emission from CO$_2^+$ in the optical regime. While \cite{Umbach_ions} identified CO$_2^+$ emission in Comet Halley within the wavelength range of 3660 to 3700\,\AA, \cite{jockers_ionsHalley_CO2+} detected CO$_2^+$ emissions within the same range, but noted an unidentified emission at 4130\,\AA. \cite{wyckoff_Halley_ions} later confirmed the 4130\,\AA~emission to be CO$_2^+$ by comparing it with a laboratory spectrum by \cite{Smyth_CO2+}, and additionally reported a CO$_2^+$ emission at around 4340\,\AA. Later, \cite{Cochran_ions_N2+_CO2+} expanded these findings by detecting CO$_2^+$ emissions in Comet Ikeya-Zhang, both within the range of 3400 to 3700\,\AA~and also at 4130\,\AA.
 
Apparently, \cite{N2plus_H2Op_g_lutz} had observed ionic emissions in Comet Bradfield 1987 but missed out on reporting a significant emission at 4130\,\AA. More recently, \cite{Opitom_R2_highres} reported the detection of CO$_2^+$ in Comet C/2016 R2 within the wavelength range of 3500 to 3660\,\AA. 
We also identified emissions of CO$_2^+$ around 4130\,\AA~in the high resolution UVES spectra of comet C/2016 R2 and in the tail spectrum of comet C/2002 T7 (LINEAR) in a current effort to build an atlas of cometary lines and still under work  \citep{Hardy_atlas}. The absence of this emission in the low-resolution spectrum of C/2016 R2 reported by \citet{kumar_16R2} (see Figure~\ref{F3_R2}), in contrast to its presence in the high resolution spectra, may be attributed to the difference in the heliocentric distance at the time of observation. The comet was at a comparatively smaller heliocentric distance during the observations reported in \cite{Opitom_R2_highres}, which may have contributed to the emergence of the emission feature. More precisely, the only lines present in that region are weak CO$_2^+$ bands that we identified in both comets C/2016 R2 and C/2002 T7 at exactly 4108.45\,\AA~ (3,8), 4109.45\,\AA~(2,5), 4121.58\,\AA~(4,7), 4123.29\,\AA~(3,6), 4138.76\,\AA~and 4139.52\,\AA, matching the laboratory CO$_2^+$ line lists of \cite{Mrozowski_CO2+_1,Mrozowski_CO2+_2} and \cite{KimCo2p_g}. At the low-resolution as of the 20F3 spectrum in the tail direction, these bands are seen to merge to give a broader feature similar to the band observed in 20F3. Moreover, the absence of any strong bands of CO$^+$ \citep{CO+_Rousselot}, CN bands (confirmed using Planetary Spectrum Generator \citep[PSG, ][]{PSG_1,PSG_2}) and C$_3$ emissions (diminished at this distance from the photocentre as seen in Figure \ref{F3_CD}) in this region, this feature around 4130\,\AA~can be attributed to CO$_2^+$. In comet C/2002 C1 (Ikeya-Zhang), which had strong emissions from CN, C$_3$ and CO$^+$, \cite{Cochran_ions_N2+_CO2+} has identified this broad feature to be CO$_2^+$. 

Hence, in addition to identifying the emissions of N$_2^+$, CO$^+$ and H$_2$O$^+$ in 20F3, we found in our low-resolution spectrum the rarely detected ionic emission corresponding to CO$_2^+$. Figure \ref{F3_CO2p} illustrates the detected emission matching with those observed in comet Ikeya-Zhang and comet Bradfield at the same position. This finding thereby clarifies that the previously unnoticed emission in the comet Bradfield also belongs to CO$_2^+$ and can be observed in low-resolution spectra. This is of some importance due to the difficulty in observing CO$_2$ from the ground. 

With the help of Equation \ref{cd}, the CO$_2^+$/CO$^+$ ratio is computed as,
 \begin{equation}\label{ion_ratio}
    \frac{CO_2^+}{CO^+}=\frac{g_{CO^+}}{g_{CO_2^+}} \frac{F_{CO_2^+}}{F_{CO^+}},
\end{equation}
where F$_{CO_2^+}$ and g$_{CO_2^+}$ are the measured band intensity of the emission and fluorescence efficiency from \cite{KimCo2p_g} (3.89$\times 10^{-15}$ ergss$^{-1}$mol$^{-1}$ at 1 au). The ratio is computed with the CO$^+$(3-0) band whose fluorescence efficiency is taken to be 1.7$\times$10$^{-14}$ ergss$^{-1}$mol$^{-1}$ at 1 au \citep{CO+_Rousselot}. We measure CO$_2^+$/CO$^+$ ratio to be 1.34$\pm$0.21, which is slightly higher than the ratio reported by \cite{Opitom_R2_highres} for comet C/2016 R2. As pointed out by \cite{Huebner_CO2_CO}, since the photodissociative ionisation of CO$_2$ could contribute significantly to CO$^+$ production in the comet coma, the ratio CO$_2^+$/CO$^+$ cannot be directly linked to CO$_2$/CO. Since reports of simultaneous measurements of CO$_2^+$ and CO$^+$ are scarce, we are limited in our comparison. A dedicated future modeling effort that quantitatively distinguishes the various sources contributing to CO$^+$ production would enable a more meaningful comparison of the derived ratios with the CO$_2$/CO values reported for other comets by \citet{feldman1997} and \cite{harrington_CO2_CO}, thereby facilitating a more robust interpretation.

\subsubsection{Higher abundance of N$_2^+$ in 20F3}\label{N2_model}
It is always challenging in low-resolution spectroscopy to fully resolve the N$_2^+$(0-0) band lying near the strong CN emission band centred at 3880\,\AA. Although \cite{F3_highres} has performed very high-resolution spectroscopy of 20F3, no ionic emissions have been reported. This may be because such high-resolution observations usually target the bright inner coma of the comet, whereas the ionic emissions are detected primarily in the direction of the ion tail, where the emission of the neutral species has dropped a lot. In these cases, long-slit low-resolution spectroscopy is advantageous as one can scan
the spatial axis to look for such emissions.

\begin{figure}[h!]
  \centering
  \includegraphics[width=0.8\linewidth]{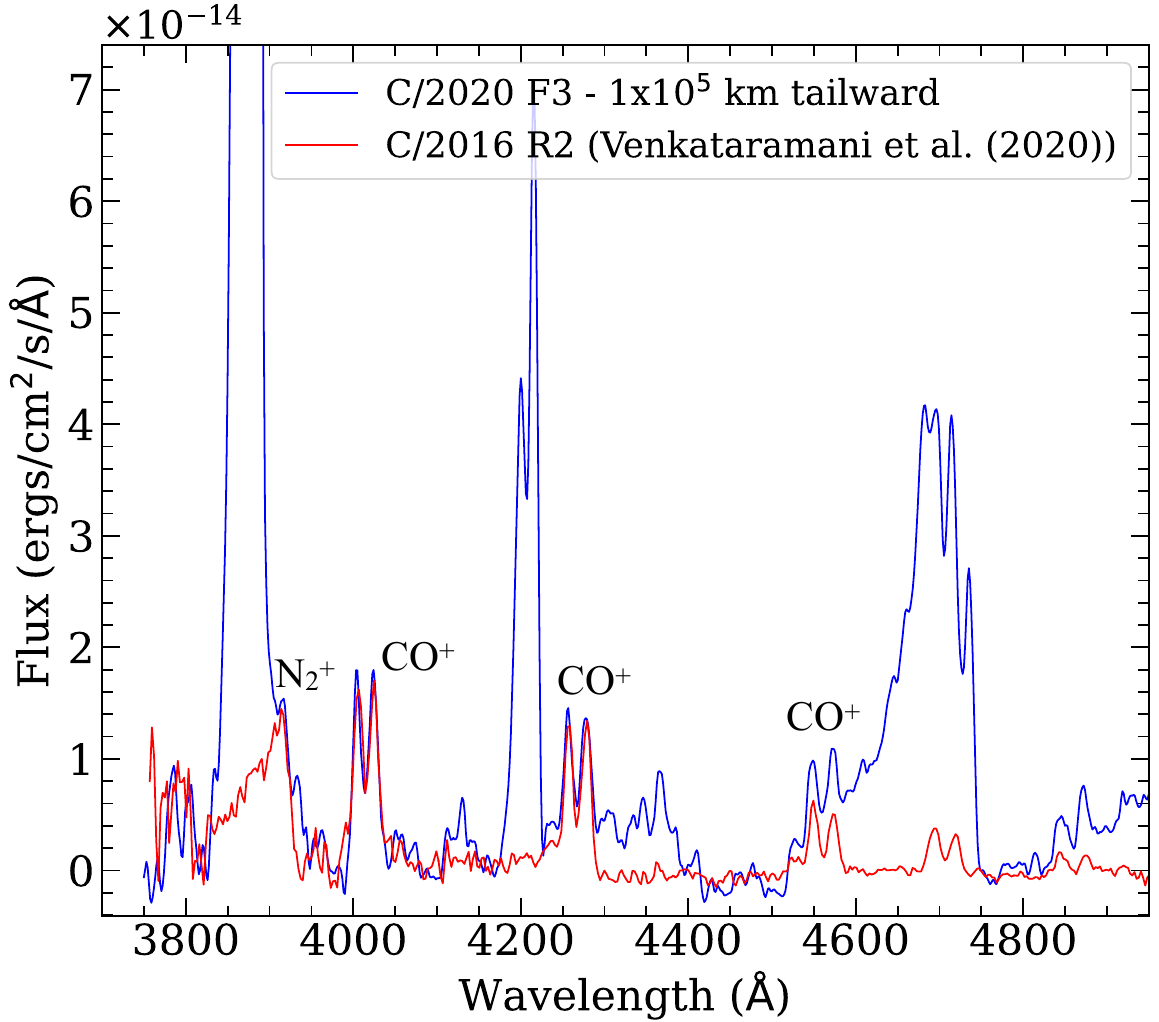}
   \caption{Comparison of the tailward spectra of 20F3 and that of comet C/2016 R2 reported in \cite{Kumar_R2}. Note that the two comets were not at all at the same distance from the Sun at the time of observation. So this is just an illustration to compare with a comet well known to be rich in N$_2^+$ and CO$^+$.}
   \label{F3_R2}
   \end{figure} 

\begin{figure}
\centering
\includegraphics[width=0.80\linewidth]{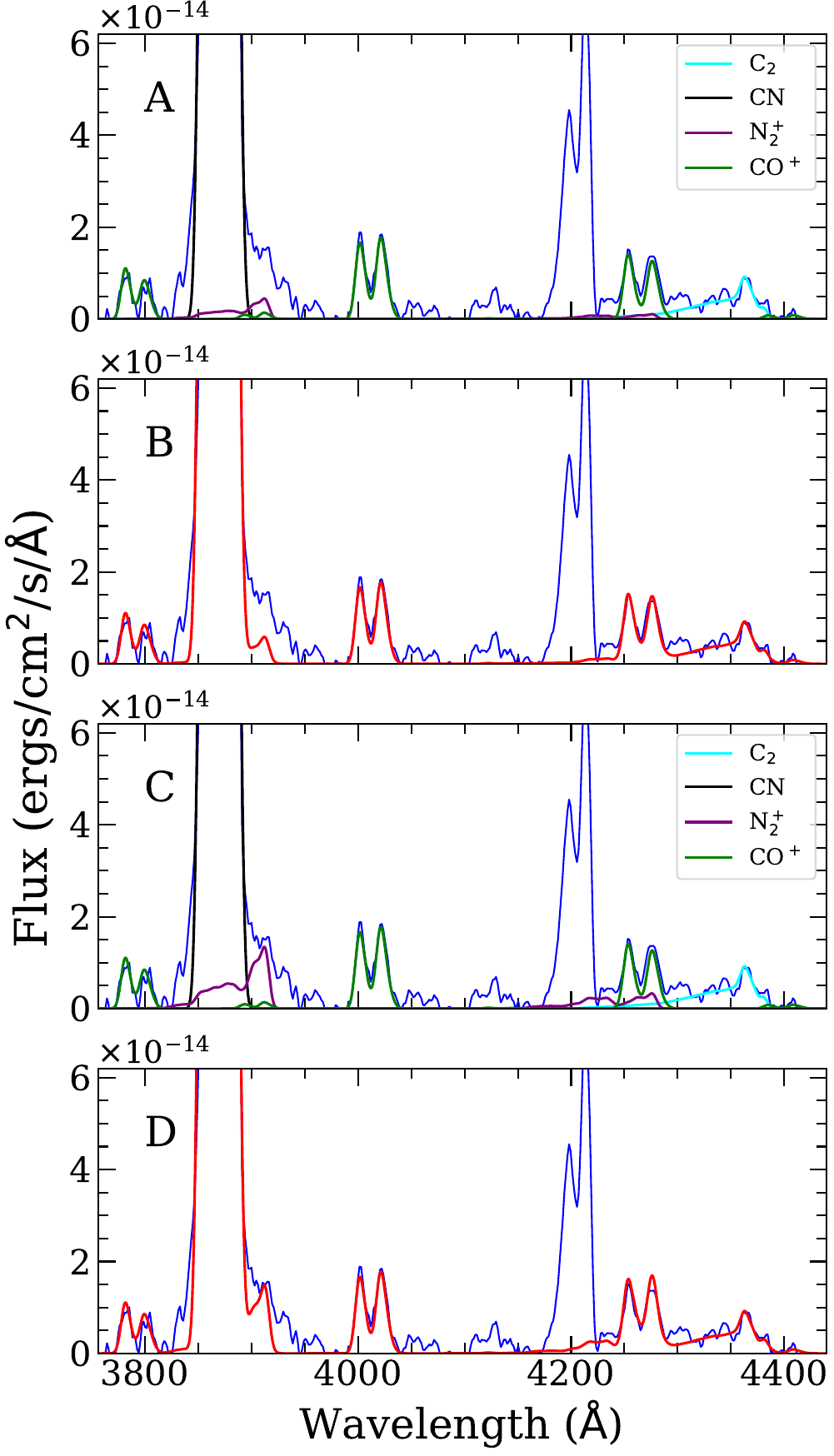}
\caption {Modeling of N$_2^+$ emission bands. Fig. A corresponds to N$_2^+$/CO$^+$=2.4$\times$10$^{-2}$ with the different species being represented separatly. Fig. B corresponds to the same ratio but only with the overall fit (red) obtained by adding the different fits for C$_2$, N$_2^+$, CO$^+$ and CN. Fig. C and D corresponds to N$_2^+$/CO$^+$=7.3$\times$10$^{-2}$. The observed spectrum is shown in blue.}
\label{fig:fit_N2plus}
\end{figure}  
   
In the absence of a high-resolution study to explore the possible higher abundance of N$_2^+$ in 20F3, the spectrum of comet C/2016 R2 (PanSTARRS) \citep{Kumar_R2}, filled with ionic emissions and without CN emission, was directly compared as shown in Figure \ref{F3_R2}. Comparing both comet spectra, we observe a similarity in the emission flux density for the CO$^+$ (3-0) and (2-0) bands as well as for the 3915\,\AA~peak of the N$_2^+$ emission. The drastic difference in the heliocentric distance of both comets at the time of observation implies differences in the fluorescence efficiency and hence the column densities. However, the good agreement in the N$_2^+$ band head between the two comets suggests the possibility of a higher N$_2^+$ flux density in 20F3 than what is measured from the visible band head.

To confirm the N$_2^+$/CO$^+$ ratio computed in Section \ref{ion_ratios}, we incorporate modelling techniques similar to those used in \cite{N2plus_rousselot}, but at lower resolution. The parameters used for this modelling are the heliocentric distance and velocity, and, because it is based on a Monte Carlo simulation, an evolutionary time of 10,000~s from an initial Boltzmann distribution, corresponding to a temperature of 80~K, has been used. We have tried to model the N$_2^+$ emission band and the spectral region surrounding it using different models for the CN, N$_2^+$, CO$^+$ and C$_2$ radicals. We used models similar to those described in \cite{anderson:2024} but convolved with a Gaussian instrument response function having an FWHM adapted for the spectrum (i.e. 8~\AA). For the C$_2$ radical, we used a model similar to that described in \cite{rousselot:2012}, calculated for the fluorescence equilibrium. Unfortunately, because the N$_2^+$ (0,0) emission bands near 3900~\AA\ are blended with both CN and CO$^+$ emission bands, it is essential to model these emission bands as well as possible.

The overall spectral flux density of the CO$^+$ emission bands can be estimated using the (3,0) (at 4000 and 4020~\AA) and (2,0) (at 4252 and 4275~\AA) emission bands (this last one being blended with the faint (5,2) emission band). In the latter band it also blended with a fainter emission band of the N$_2^+$ (0,1) emission band and C$_2$ emission bands. A good fit of the (3,0) and (2,0) CO$^+$ emission bands, including the N$_2^+$ (0,1) band, can be achieved this a ratio N$_2^+$/CO$^+$=2.4$\times$10$^{-2}$. With such a ratio, which is probably the lowest possible value, the emission band N$_2^+$ (0,0) (the brightest) near the CN bright emission band cannot account for all the observed flux density at 3914~\AA. An upper limit for the N$_2^+$/CO$^+$ ratio can be estimated by fitting the flux density of the N$_2^+$ (0,0) band to the observation spectrum. In this case, we get N$_2^+$/CO$^+$=7.3$\times$10$^{-2}$. With such a ratio, the fit of CO$^+$ (2,0) + N$_2^+$ (0,1) + C$_2$ at 4275~\AA\ appears overestimated compared to the observational spectrum. In this case, there are still a few unexplained emission lines near the N$_2^+$ emission bands, probably due to other species. These species are probably also responsible for part of the flux observed at 3914~\AA. 

Fig. \ref{fig:fit_N2plus} presents the different fits. Taking into account both the findings and concluding that the ratio is the mean of the computed lower and upper limits, we report the N$_2^+$/CO$^+$ to be $(4.8\pm2.4)\times10^{-2}$. Similar to the interpretation provided by \cite{Opitom_R2_highres}, since CO$_2^+$ has been detected in comet 20F3 and the photodissociative ionisation of CO$_2$ can also produce CO$^+$, the computed N$_2$/CO in this work should be considered as a lower limit.

\section{Conclusions}
In this work, we present the long-slit low-resolution optical spectroscopic and photometric observation of the bright comet of 2020, C/2020 F3 (NEOWISE), using the 2-m HCT and TRAPPIST-North telescope. Spatial spectrum analysis unveils clear emissions from the ionic species N$_2^+$, CO$^+$, CO$_2^+$ and H$_2$O$^+$ in addition to the regular CN, C$_2$, C$_3$, NH$_2$ and [OI] emissions. Using these observational results, we arrive at the following conclusions:
   \begin{enumerate}
      \item The computed production rate ratios, despite showing variations across the observation, classify the comet as gas-rich, having a typical carbon-chain composition.
      \item The water production rate computed using the spectral flux density of the forbidden oxygen line [OI] at 6300\,\AA~is in agreement with those reported by space-based and ground-based measurements. These values may be regarded as reliable upper limits on the water production rates.
      \item The average nucleus rotation period was found to be 7.28 $\pm$ 0.79 hours from the observation of the CN jets, and the gas velocity was found to be 2.40$\pm$0.25 km/s.
      \item The average N$_2$/CO for this comet, computed from the spectral observation, was found to be $(3.0\pm1.0)\times10^{-2}$, which is uniform within the measurement errors, from $\sim$ 40 000 km to $\sim$100 000 km from the photo centre. 
      \item Taking into account the similarity of flux density for CO$^+$ and N$_2^+$ in both 20F3 and C/2016 R2, fluorescence modelling techniques were used to refine the ratio. With this we estimate the N$_2$/CO to be $(4.8\pm2.4)\times10^{-2}$. This ratio is considered as a lower limit accounting for the detection of CO$_2^+$ in the comet.
      \item In addition to the usual ionic emissions of N$_2^+$, CO$^+$ and H$_2$O$^+$, an emission at 4130\,\AA~corresponding to CO$_2^+$ was identified. Hence, 20F3 joins the list of great comets with identified CO$_2^+$ emission and shows that this species can be detected from low resolution spectra.
      \item The CO$_2^+$/CO$^+$ ratio was computed to be 1.34$\pm$0.21. Further modelling efforts are necessary to directly link the observed ratio with the known CO$_2$/CO ratio of other comets.
      \item The high CO$_2^+$/CO$^+$ ratio of 1.34 observed in C/2020 F3, combined with a relatively low N$_2^+$/CO$^+$ ratio of 0.048, suggests that the comet probably formed in a region of the solar nebula where temperatures were cold enough for CO$_2$ to condense efficiently, but too warm to trap significant N$_2$. This is consistent with formation beyond the CO$_2$ ice line ($\sim$70~K), but not far enough to retain more volatile species such as N$_2$, indicating a likely origin in the moderately cold mid-outer nebula ($\sim$50~K-70~K). Extending this type of analysis to other comets is important for refining our interpretation of the results.
   \end{enumerate}

\begin{acknowledgements}
We thank the staff of the Indian Astronomical Observatory, Hanle and Centre For Research \& Education in Science \& Technology, Hoskote, that made these observations possible. The facilities at IAO and CREST are operated by the Indian Institute of Astrophysics, Bangalore. Work at Physical Research Laboratory is supported by the Department of Space, Govt. of India. \\
This publication makes use of data products from the TRAPPIST project, under the scientific direction of Prof. Emmanuel Jehin, Director of Research at the Belgian National Fund for Scientific Research (F.R.S.-FNRS). M.V.D is supported by the FNRS (Fond National pour la Recherche Scientifique; Belgium) throught its FRIA PhD grant. TRAPPIST is funded by F.R.S.-FNRS under grant PDR T.0120.21, and TRAPPIST-North is funded by the University of Liège in collaboration with Cadi Ayyad University of Marrakech. The authors thank NASA, D. Schleicher, and the Lowell Observatory for the loan of a set of HB comet filters.

A.K. acknowledges support from the Wallonia-Brussels International (WBI) grant. This work is a result of the bilateral Belgo-Indian projects on Precision Astronomical Spectroscopy for Stellar and Solar System bodies, BIPASS, funded by the Belgian Federal Science Policy Office (BELSPO, Govt. of Belgium; BL/33/IN22\_BIPASS) and the International Division, Department of Science and Technology (DST, Govt. of India; DST/INT/BELG/P-01/2021 (G)).
\end{acknowledgements}

\bibliographystyle{aa} 
\bibliography{reference.bib} 

\begin{thebibliography}{84}
\expandafter\ifx\csname natexlab\endcsname\relax\def\natexlab#1{#1}\fi

\bibitem[{A'Hearn {et~al.}(1995)A'Hearn, Millis, Schleicher, Osip, \&
  Birch}]{Ahearn_85}
A'Hearn, M.~F., Millis, R.~C., Schleicher, D.~O., Osip, D.~J., \& Birch, P.~V.
  1995, Icarus, 118, 223

\bibitem[{{A'Hearn} {et~al.}(1984){A'Hearn}, {Schleicher}, {Millis}, {Feldman},
  \& {Thompson}}]{A'Hearn1984}
{A'Hearn}, M.~F., {Schleicher}, D.~G., {Millis}, R.~L., {Feldman}, P.~D., \&
  {Thompson}, D.~T. 1984, The Astronomical Journal, 89, 579

\bibitem[{{Anderson} {et~al.}(2024){Anderson}, {Rousselot}, {Jehin},
  {Noyelles}, {Manfroid}, {Hardy}, \& {Robert}}]{anderson:2024}
{Anderson}, S.~E., {Rousselot}, P., {Jehin}, E., {et~al.} 2024, \aap, 692, A1

\bibitem[{{Anderson} {et~al.}(2023){Anderson}, {Rousselot}, {Noyelles},
  {Jehin}, \& {Mousis}}]{N2_CO_orbital}
{Anderson}, S.~E., {Rousselot}, P., {Noyelles}, B., {Jehin}, E., \& {Mousis},
  O. 2023, \mnras, 524, 5182

\bibitem[{{Aravind} {et~al.}(2021){Aravind}, {Ganesh}, {Venkataramani}, {Sahu},
  {Angchuk}, {Sivarani}, \& {Unni}}]{borisov_aravind}
{Aravind}, K., {Ganesh}, S., {Venkataramani}, K., {et~al.} 2021, \mnras, 502,
  3491

\bibitem[{Aravind {et~al.}(2022)Aravind, Halder, Ganesh, Sahu, Serra-Ricart,
  Chambó, Angchuk, \& Sivarani}]{156P_aravind}
Aravind, K., Halder, P., Ganesh, S., {et~al.} 2022, Icarus, 383, 115042

\bibitem[{{Biermann} {et~al.}(1982){Biermann}, {Giguere}, \&
  {Huebner}}]{N2_H2O_modelling}
{Biermann}, L., {Giguere}, P.~T., \& {Huebner}, W.~F. 1982, \aap, 108, 221

\bibitem[{{Biver} {et~al.}(2018){Biver}, {Bockel{\'e}e-Morvan}, {Paubert},
  {Moreno}, {Crovisier}, {Boissier}, {Bertrand}, {Boussier}, {Kugel}, {McKay},
  {Dello Russo}, \& {DiSanti}}]{Biver_C2016R2}
{Biver}, N., {Bockel{\'e}e-Morvan}, D., {Paubert}, G., {et~al.} 2018, \aap,
  619, A127

\bibitem[{{Bodewits} {et~al.}(2019){Bodewits}, {Orsz{\'a}gh}, {Noonan},
  {{\v{D}}urian}, \& {Matej{\v{c}}{\'\i}k}}]{bodewits_H2O+}
{Bodewits}, D., {Orsz{\'a}gh}, J., {Noonan}, J., {{\v{D}}urian}, M., \&
  {Matej{\v{c}}{\'\i}k}, {\v{S}}. 2019, \apj, 885, 167

\bibitem[{{Bradley} {et~al.}(2016){Bradley}, {Sipocz}, {Robitaille},
  {Tollerud}, {Deil}, {Vin{\'\i}cius}, {Barbary}, {G{\"u}nther}, {Bostroem},
  {Droettboom}, {Bray}, {Bratholm}, {Pickering}, {Craig}, {Pascual}, {Greco},
  {Donath}, {Kerzendorf}, {Littlefair}, {Barentsen}, {D'Eugenio}, \&
  {Weaver}}]{photutils}
{Bradley}, L., {Sipocz}, B., {Robitaille}, T., {et~al.} 2016, {Photutils:
  Photometry tools}, Astrophysics Source Code Library, record ascl:1609.011

\bibitem[{{Cambianica} {et~al.}(2021){Cambianica}, {Cremonese}, {Munaretto},
  {Capria}, {Fulle}, {Boschin}, {Di Fabrizio}, \& {Harutyunyan}}]{F3_highres}
{Cambianica}, P., {Cremonese}, G., {Munaretto}, G., {et~al.} 2021, \aap, 656,
  A160

\bibitem[{{Cochran}(1985)}]{Cochran_scalelength}
{Cochran}, A.~L. 1985, \aj, 90, 2609

\bibitem[{{Cochran}(2002)}]{Cochran_ions_N2+_CO2+}
{Cochran}, A.~L. 2002, \apjl, 576, L165

\bibitem[{{Cochran} {et~al.}(2012){Cochran}, {Barker}, \&
  {Gray}}]{cochran_30years}
{Cochran}, A.~L., {Barker}, E.~S., \& {Gray}, C.~L. 2012, \icarus, 218, 144

\bibitem[{{Cochran} {et~al.}(2000){Cochran}, {Cochran}, \&
  {Barker}}]{cochran_N2+_CO+_many}
{Cochran}, A.~L., {Cochran}, W.~D., \& {Barker}, E.~S. 2000, \icarus, 146, 583

\bibitem[{{Cochran} \& {McKay}(2018)}]{Cochran_C2016R2}
{Cochran}, A.~L. \& {McKay}, A.~J. 2018, \apjl, 854, L10

\bibitem[{{Cochran} {et~al.}(2020){Cochran}, {Nelson}, {McKay}, {MacQueen},
  {Cochran}, \& {Endl}}]{Cochran2020DPS}
{Cochran}, A.~L., {Nelson}, T., {McKay}, A.~J., {et~al.} 2020, in AAS/Division
  for Planetary Sciences Meeting Abstracts, Vol.~52, AAS/Division for Planetary
  Sciences Meeting Abstracts, 111.03

\bibitem[{{Combi} {et~al.}(2021){Combi}, {M{\"a}kinen}, {Bertaux},
  {Qu{\'e}merais}, \& {Ferron}}]{C2020F3_combi_waterprod}
{Combi}, M.~R., {M{\"a}kinen}, T., {Bertaux}, J.~L., {Qu{\'e}merais}, E., \&
  {Ferron}, S. 2021, \apjl, 907, L38

\bibitem[{{Dello Russo} {et~al.}(2016){Dello Russo}, {Vervack}, {Kawakita},
  {Cochran}, {McKay}, {Harris}, {Weaver}, {Lisse}, {DiSanti}, {Kobayashi},
  {Biver}, {Bockel{\'e}e-Morvan}, {Crovisier}, {Opitom}, \&
  {Jehin}}]{dellorusso_OH_H2O}
{Dello Russo}, N., {Vervack}, R.~J., {Kawakita}, H., {et~al.} 2016, \icarus,
  266, 152

\bibitem[{{Drahus} {et~al.}(2020){Drahus}, {Guzik}, {Stephens}, {Howell},
  {Zola}, {Sabat}, \& {Reichart}}]{Drahus2020ATel}
{Drahus}, M., {Guzik}, P., {Stephens}, A., {et~al.} 2020, The Astronomer's
  Telegram, 13945, 1

\bibitem[{{Faggi} {et~al.}(2021){Faggi}, {Lippi}, {Camarca}, {Buzard},
  {Villanueva}, {Doppmann}, {Blake}, \& {Mumma}}]{F3_heterogenous}
{Faggi}, S., {Lippi}, M., {Camarca}, M., {et~al.} 2021, \aj, 162, 178

\bibitem[{{Farnham} {et~al.}(2000){Farnham}, {Schleicher}, \&
  {A'Hearn}}]{Farnham2000}
{Farnham}, T.~L., {Schleicher}, D.~G., \& {A'Hearn}, M.~F. 2000, Icarus, 147,
  180

\bibitem[{{Fegley} \& {Prinn}(1989)}]{solar_nebula_N2_CO}
{Fegley}, Bruce, J. \& {Prinn}, R.~G. 1989, in The Formation and Evolution of
  Planetary Systems, ed. H.~A. {Weaver} \& L.~{Danly}, 171--205

\bibitem[{{Feldman} {et~al.}(1986){Feldman}, {A'Hearn}, {Festou}, {McFadden},
  {Weaver}, \& {Woods}}]{Feldman_CO2p}
{Feldman}, P.~D., {A'Hearn}, M.~F., {Festou}, M.~C., {et~al.} 1986, \nat, 324,
  433

\bibitem[{{Feldman} {et~al.}(2004){Feldman}, {Cochran}, \&
  {Combi}}]{N2_feldman}
{Feldman}, P.~D., {Cochran}, A.~L., \& {Combi}, M.~R. 2004, in Comets II, ed.
  M.~C. {Festou}, H.~U. {Keller}, \& H.~A. {Weaver}, 425

\bibitem[{{Feldman} {et~al.}(1997){Feldman}, {Festou}, {Tozzi}, \&
  {Weaver}}]{feldman1997}
{Feldman}, P.~D., {Festou}, M.~C., {Tozzi}, P., \& {Weaver}, H.~A. 1997, \apj,
  475, 829

\bibitem[{{Festou} {et~al.}(1982){Festou}, {Feldman}, \&
  {Weaver}}]{bradfield_seargent}
{Festou}, M.~C., {Feldman}, P.~D., \& {Weaver}, H.~A. 1982, \apj, 256, 331

\bibitem[{{Fink}(2009)}]{Fink_comet_survey_2009}
{Fink}, U. 2009, \icarus, 201, 311

\bibitem[{{Fink} \& {Combi}(2004)}]{Fink_46P_scalelength}
{Fink}, U. \& {Combi}, M.~R. 2004, \planss, 52, 573

\bibitem[{{Fink} {et~al.}(1991){Fink}, {Combi}, \&
  {Disanti}}]{Fink_scalelength}
{Fink}, U., {Combi}, M.~R., \& {Disanti}, M.~A. 1991, \apj, 383, 356

\bibitem[{{Fink} \& {Hicks}(1996)}]{Fink1996}
{Fink}, U. \& {Hicks}, M.~D. 1996, \apj, 459, 729

\bibitem[{{Hardy} {et~al.}(2023){Hardy}, {Jehin}, {Rousselot},
  {Hutsem{\'e}kers}, \& {Manfroid}}]{Hardy_atlas}
{Hardy}, P., {Jehin}, E., {Rousselot}, P., {Hutsem{\'e}kers}, D., \&
  {Manfroid}, J. 2023, in LPI Contributions, Vol. 2851, LPI Contributions, 2148

\bibitem[{{Harrington Pinto} {et~al.}(2022){Harrington Pinto}, {Womack},
  {Fernandez}, \& {Bauer}}]{harrington_CO2_CO}
{Harrington Pinto}, O., {Womack}, M., {Fernandez}, Y., \& {Bauer}, J. 2022,
  \psj, 3, 247

\bibitem[{{Haser}(1957)}]{haser}
{Haser}, L. 1957, Bulletin de la Societe Royale des Sciences de Liege, 43, 740

\bibitem[{{Haser} {et~al.}(2020){Haser}, {Oset}, \&
  {Bodewits}}]{bodewits_haser}
{Haser}, L., {Oset}, S., \& {Bodewits}, D. 2020, \psj, 1, 83

\bibitem[{{Huebner} \& {Giguere}(1980)}]{Huebner_CO2_CO}
{Huebner}, W.~F. \& {Giguere}, P.~T. 1980, \apj, 238, 753

\bibitem[{{Ivanova} {et~al.}(2021){Ivanova}, {Rosenbush}, {Luk'yanyk},
  {Kolokolova}, {Kleshchonok}, {Kiselev}, {Afanasiev}, \& {Ren{\'e}e
  Kirk}}]{ivanova_2021}
{Ivanova}, O., {Rosenbush}, V., {Luk'yanyk}, I., {et~al.} 2021, \aap, 651, A29

\bibitem[{{Jehin} {et~al.}(2011){Jehin}, {Gillon}, {Queloz}, {Magain},
  {Manfroid}, {Chantry}, {Lendl}, {Hutsem{\'e}kers}, \& {Udry}}]{Jehin2011}
{Jehin}, E., {Gillon}, M., {Queloz}, D., {et~al.} 2011, The Messenger, 145, 2

\bibitem[{{Jockers} {et~al.}(1987){Jockers}, {Rosenbauer}, {Geyer}, \&
  {Haenel}}]{jockers_ionsHalley_CO2+}
{Jockers}, K., {Rosenbauer}, H., {Geyer}, E.~H., \& {Haenel}, A. 1987, \aap,
  187, 256

\bibitem[{{Kawakita} \& {Watanabe}(2002)}]{Kawakita_ions_unidentified}
{Kawakita}, H. \& {Watanabe}, J.-i. 2002, \apjl, 574, L183

\bibitem[{{Kim}(1999)}]{KimCo2p_g}
{Kim}, S.~J. 1999, Earth, Planets and Space, 51, 139

\bibitem[{{Korsun} {et~al.}(2006){Korsun}, {Ivanova}, \&
  {Afanasiev}}]{Korsun_C2002VQ94}
{Korsun}, P.~P., {Ivanova}, O.~V., \& {Afanasiev}, V.~L. 2006, \aap, 459, 977

\bibitem[{{Krishnakumar} {et~al.}(2020){Krishnakumar}, {Angchuk},
  {Venkataramani}, {Ganesh}, {Sahu}, \& {Sivarani}}]{atel_aravind}
{Krishnakumar}, A., {Angchuk}, D., {Venkataramani}, K., {et~al.} 2020, The
  Astronomer's Telegram, 13897, 1

\bibitem[{{Langland-Shula} \& {Smith}(2011)}]{langland-shula}
{Langland-Shula}, L.~E. \& {Smith}, G.~H. 2011, \icarus, 213, 280

\bibitem[{{Lew}(1976)}]{Lew_h2op}
{Lew}, H. 1976, Canadian Journal of Physics, 54, 2028

\bibitem[{{Lin} {et~al.}(2020){Lin}, {Wang}, {Ip}, {Huang}, {Lin}, {Hsiao},
  {Hou}, \& {Lin}}]{Lin2020ATel}
{Lin}, Z.-Y., {Wang}, C., {Ip}, W.-H., {et~al.} 2020, The Astronomer's
  Telegram, 13886, 1

\bibitem[{{Lodders} {et~al.}(2009){Lodders}, {Palme}, \&
  {Gail}}]{lodders_solar_nebula}
{Lodders}, K., {Palme}, H., \& {Gail}, H.~P. 2009, Landolt B{\"o}rnstein, 4B,
  712

\bibitem[{{Lutz} {et~al.}(1993){Lutz}, {Womack}, \&
  {Wagner}}]{N2plus_H2Op_g_lutz}
{Lutz}, B.~L., {Womack}, M., \& {Wagner}, R.~M. 1993, \apj, 407, 402

\bibitem[{{Mainzer} {et~al.}(2014){Mainzer}, {Bauer}, {Cutri}, {Grav},
  {Masiero}, {Beck}, {Clarkson}, {Conrow}, {Dailey}, {Eisenhardt}, {Fabinsky},
  {Fajardo-Acosta}, {Fowler}, {Gelino}, {Grillmair}, {Heinrichsen}, {Kendall},
  {Kirkpatrick}, {Liu}, {Masci}, {McCallon}, {Nugent}, {Papin}, {Rice},
  {Royer}, {Ryan}, {Sevilla}, {Sonnett}, {Stevenson}, {Thompson}, {Wheelock},
  {Wiemer}, {Wittman}, {Wright}, \& {Yan}}]{NEOWISE}
{Mainzer}, A., {Bauer}, J., {Cutri}, R.~M., {et~al.} 2014, \apj, 792, 30

\bibitem[{{Manzini} {et~al.}(2021){Manzini}, {Oldani}, {Ochner}, {Barbotin},
  {Bedin}, {Behrend}, \& {Fardelli}}]{manzini_rotation}
{Manzini}, F., {Oldani}, V., {Ochner}, P., {et~al.} 2021, \mnras, 506, 6195

\bibitem[{{Marcus}(2007)}]{Marcus2007}
{Marcus}, J.~N. 2007, International Comet Quarterly, 29, 39

\bibitem[{{McKay} {et~al.}(2012){McKay}, {Chanover}, {Morgenthaler}, {Cochran},
  {Harris}, \& {Russo}}]{Mckay_2012_oxygen}
{McKay}, A.~J., {Chanover}, N.~J., {Morgenthaler}, J.~P., {et~al.} 2012,
  \icarus, 220, 277

\bibitem[{{McKay} {et~al.}(2020){McKay}, {Cochran}, {Dello Russo}, \&
  {DiSanti}}]{Mckay_borisov_oxygen}
{McKay}, A.~J., {Cochran}, A.~L., {Dello Russo}, N., \& {DiSanti}, M.~A. 2020,
  \apjl, 889, L10

\bibitem[{{Mommert} {et~al.}(2019){Mommert}, {Kelley}, {de Val-Borro}, {Li},
  {Guzman}, {Sipo{\`I}cz}, {D{\`I}urech}, {Granvik}, {Grundy}, {Moskovitz},
  {Penttil{\"a}}, \& {Samarasinha}}]{sbpy}
{Mommert}, M., {Kelley}, M. S.~P., {de Val-Borro}, M., {et~al.} 2019, {sbpy:
  Small-body planetary astronomy}, Astrophysics Source Code Library, record
  ascl:1907.014

\bibitem[{{Morgenthaler} {et~al.}(2007){Morgenthaler}, {Harris}, \&
  {Combi}}]{Morgenthaler_2007}
{Morgenthaler}, J.~P., {Harris}, W.~M., \& {Combi}, M.~R. 2007, \apj, 657, 1162

\bibitem[{{Morgenthaler} {et~al.}(2001){Morgenthaler}, {Harris}, {Scherb},
  {Anderson}, {Oliversen}, {Doane}, {Combi}, {Marconi}, \&
  {Smyth}}]{Morgenthaler_2001}
{Morgenthaler}, J.~P., {Harris}, W.~M., {Scherb}, F., {et~al.} 2001, \apj, 563,
  451

\bibitem[{{Moulane} {et~al.}(2023){Moulane}, {Jehin}, {Manfroid},
  {Hutsem{\'e}kers}, {Opitom}, {Shinnaka}, {Bodewits}, {Benkhaldoun}, {Jabiri},
  {Hmiddouch}, {Vander Donckt}, {Pozuelos}, \& {Yang}}]{moulane_46p}
{Moulane}, Y., {Jehin}, E., {Manfroid}, J., {et~al.} 2023, \aap, 670, A159

\bibitem[{{Moulane} {et~al.}(2018){Moulane}, {Jehin}, {Opitom}, {Pozuelos},
  {Manfroid}, {Benkhaldoun}, {Daassou}, \& {Gillon}}]{Moulane2018}
{Moulane}, Y., {Jehin}, E., {Opitom}, C., {et~al.} 2018, \aap, 619, A156

\bibitem[{{Mrozowski}(1947{\natexlab{a}})}]{Mrozowski_CO2+_2}
{Mrozowski}, S. 1947{\natexlab{a}}, Physical Review, 72, 682

\bibitem[{{Mrozowski}(1947{\natexlab{b}})}]{Mrozowski_CO2+_1}
{Mrozowski}, S. 1947{\natexlab{b}}, Physical Review, 72, 691

\bibitem[{{Opitom} {et~al.}(2020){Opitom}, {Guilbert-Lepoutre}, {Besse},
  {Yang}, \& {Snodgrass}}]{67P_muse_oxygen}
{Opitom}, C., {Guilbert-Lepoutre}, A., {Besse}, S., {Yang}, B., \& {Snodgrass},
  C. 2020, \aap, 644, A143

\bibitem[{{Opitom} {et~al.}(2019){Opitom}, {Hutsem{\'e}kers}, {Jehin},
  {Rousselot}, {Pozuelos}, {Manfroid}, {Moulane}, {Gillon}, \&
  {Benkhaldoun}}]{Opitom_R2_highres}
{Opitom}, C., {Hutsem{\'e}kers}, D., {Jehin}, E., {et~al.} 2019, \aap, 624, A64

\bibitem[{{Opitom} {et~al.}(2015){Opitom}, {Jehin}, {Manfroid},
  {Hutsem{\'e}kers}, {Gillon}, \& {Magain}}]{opitom_lemmon}
{Opitom}, C., {Jehin}, E., {Manfroid}, J., {et~al.} 2015, \aap, 574, A38

\bibitem[{{Owen} \& {Bar-Nun}(1995)}]{owen_N2_CO}
{Owen}, T. \& {Bar-Nun}, A. 1995, \icarus, 116, 215

\bibitem[{{Rousselot} {et~al.}(2022){Rousselot}, {Anderson}, {Alijah},
  {Noyelles}, {Opitom}, {Jehin}, {Hutsem{\'e}kers}, \&
  {Manfroid}}]{N2plus_rousselot}
{Rousselot}, P., {Anderson}, S.~E., {Alijah}, A., {et~al.} 2022, \aap, 661,
  A131

\bibitem[{{Rousselot} {et~al.}(2024){Rousselot}, {Jehin}, {Hutsem{\'e}kers},
  {Opitom}, {Manfroid}, \& {Hardy}}]{CO+_Rousselot}
{Rousselot}, P., {Jehin}, E., {Hutsem{\'e}kers}, D., {et~al.} 2024, \aap, 683,
  A50

\bibitem[{{Rousselot} {et~al.}(2012){Rousselot}, {Jehin}, {Manfroid}, \&
  {Hutsem{\'e}kers}}]{rousselot:2012}
{Rousselot}, P., {Jehin}, E., {Manfroid}, J., \& {Hutsem{\'e}kers}, D. 2012,
  \aap, 545, A24

\bibitem[{{Rubin} {et~al.}(2015){Rubin}, {Altwegg}, {Balsiger}, {Bar-Nun},
  {Berthelier}, {Bieler}, {Bochsler}, {Briois}, {Calmonte}, {Combi}, {De
  Keyser}, {Dhooghe}, {Eberhardt}, {Fiethe}, {Fuselier}, {Gasc}, {Gombosi},
  {Hansen}, {H{\"a}ssig}, {J{\"a}ckel}, {Kopp}, {Korth}, {Le Roy}, {Mall},
  {Marty}, {Mousis}, {Owen}, {R{\`e}me}, {S{\'e}mon}, {Tzou}, {Waite}, \&
  {Wurz}}]{rubin_solar_nebula}
{Rubin}, M., {Altwegg}, K., {Balsiger}, H., {et~al.} 2015, Science, 348, 232

\bibitem[{{Schleicher}(2010)}]{schleicher_CN_2010}
{Schleicher}, D.~G. 2010, \aj, 140, 973

\bibitem[{{Schleicher} {et~al.}(1998){Schleicher}, {Millis}, \&
  {Birch}}]{Schleicher1998}
{Schleicher}, D.~G., {Millis}, R.~L., \& {Birch}, P.~V. 1998, \icarus, 132, 397

\bibitem[{{Schultz} {et~al.}(1992){Schultz}, {Li}, {Scherb}, \&
  {Roesler}}]{schultz_1992_oxygen}
{Schultz}, D., {Li}, G.~S.~H., {Scherb}, F., \& {Roesler}, F.~L. 1992, \icarus,
  96, 190

\bibitem[{{Sekanina} \& {Larson}(1984)}]{larson_sekanina}
{Sekanina}, Z. \& {Larson}, S.~M. 1984, \aj, 89, 1408

\bibitem[{{Smyth}(1931)}]{Smyth_CO2+}
{Smyth}, H.~D. 1931, Physical Review, 38, 2000

\bibitem[{{Tseng} {et~al.}(2007){Tseng}, {Bockel{\'e}e-Morvan}, {Crovisier},
  {Colom}, \& {Ip}}]{Tseng_velocity}
{Tseng}, W.~L., {Bockel{\'e}e-Morvan}, D., {Crovisier}, J., {Colom}, P., \&
  {Ip}, W.~H. 2007, \aap, 467, 729

\bibitem[{{Umbach} {et~al.}(1998){Umbach}, {Jockers}, \& {Geyer}}]{Umbach_ions}
{Umbach}, R., {Jockers}, K., \& {Geyer}, E.~H. 1998, \aaps, 127, 479

\bibitem[{{Venkataramani} {et~al.}(2020{\natexlab{a}}){Venkataramani},
  {Ganesh}, \& {Baliyan}}]{Kumar_R2}
{Venkataramani}, K., {Ganesh}, S., \& {Baliyan}, K.~S. 2020{\natexlab{a}},
  \mnras, 495, 3559

\bibitem[{{Venkataramani} {et~al.}(2020{\natexlab{b}}){Venkataramani},
  {Ganesh}, \& {Baliyan}}]{kumar_16R2}
{Venkataramani}, K., {Ganesh}, S., \& {Baliyan}, K.~S. 2020{\natexlab{b}},
  \mnras, 495, 3559

\bibitem[{{Villanueva} {et~al.}(2022){Villanueva}, {Liuzzi}, {Faggi},
  {Protopapa}, {Kofman}, {Fauchez}, {Stone}, \& {Mandell}}]{PSG_2}
{Villanueva}, G.~L., {Liuzzi}, G., {Faggi}, S., {et~al.} 2022, {Fundamentals of
  the Planetary Spectrum Generator}

\bibitem[{{Villanueva} {et~al.}(2018){Villanueva}, {Smith}, {Protopapa},
  {Faggi}, \& {Mandell}}]{PSG_1}
{Villanueva}, G.~L., {Smith}, M.~D., {Protopapa}, S., {Faggi}, S., \&
  {Mandell}, A.~M. 2018, \jqsrt, 217, 86

\bibitem[{{Weaver} {et~al.}(1993){Weaver}, {Feldman}, {McPhate}, {A'Hearn},
  {Arpigny}, \& {Smith}}]{weaver}
{Weaver}, H.~A., {Feldman}, P.~D., {McPhate}, J.~B., {et~al.} 1993, in LPI
  Contributions, Vol. 810, Asteroids, Comets, Meteors 1993, ed. {LPI Editorial
  Board}, 309

\bibitem[{{Wyckoff} {et~al.}(1999){Wyckoff}, {Heyd}, \& {Fox}}]{wyckoff_unid}
{Wyckoff}, S., {Heyd}, R.~S., \& {Fox}, R. 1999, \apjl, 512, L73

\bibitem[{{Wyckoff} \& {Wehinger}(1976)}]{wyckoff_ionsincometails}
{Wyckoff}, S. \& {Wehinger}, P.~A. 1976, \apj, 204, 604

\bibitem[{{Wyckoff} {et~al.}(1986){Wyckoff}, {Wehinger}, {Spinrad}, \&
  {Belton}}]{wyckoff_Halley_ions}
{Wyckoff}, S., {Wehinger}, P.~A., {Spinrad}, H., \& {Belton}, M.~J.~S. 1986, in
  ESA Special Publication, Vol. 250, ESLAB Symposium on the Exploration of
  Halley's Comet, ed. B.~{Battrick}, E.~J. {Rolfe}, \& R.~{Reinhard}, 311--316

\bibitem[{{Ye} {et~al.}(2020){Ye}, {Zhang}, {Brewer}, {Knight}, \&
  {Kelley}}]{Ye2020DPS}
{Ye}, Q., {Zhang}, Q., {Brewer}, J., {Knight}, M., \& {Kelley}, M. 2020, in
  AAS/Division for Planetary Sciences Meeting Abstracts, Vol.~52, AAS/Division
  for Planetary Sciences Meeting Abstracts, 111.02

\end{thebibliography}

\end{document}